\documentclass{PoS}

\usepackage{wrapfig}
\usepackage{fixltx2e}

\usepackage{tikz}
\definecolor{valkoinen}{rgb}{1 1 1}
\definecolor{hysininen}{rgb}{0 0 0}
\definecolor{hyharmaa}{rgb}{0.5 0.5 0.5}
\definecolor{mloranssi}{rgb}{1 1 1}
\definecolor{vaaleaoranssi}{rgb}{1 1 1}
\definecolor{vaalvihrea}{rgb}{0.2 0.9 0.1}
\definecolor{omaviol}{rgb}{0.7 0.1 0.7}
\definecolor{omavihrea}{rgb}{0.1 0.3 0.1}
\definecolor{omapun}{rgb}{0.5 0.0 0.15}
\definecolor{omasin}{rgb}{0.0 0.15 0.5}
\definecolor{omaor}{rgb}{0.45 0.3 0}

\newcommand{\ud}{\, \mathrm{d}}
\newcommand{\nc}{{N_\mathrm{c}}}

\newcommand{\as}{\alpha_{\mathrm{s}}}

\newcommand{\tr}{\, \mathrm{Tr} \, }

\newcommand{\ra}{R_A}
\newcommand{\rp}{R_p}

\newcommand{\ampli}{{\mathcal{N}}}

\newcommand{\nr}[1]{(\ref{#1})}

\newcommand{\ptt}{{p_T}}
\newcommand{\ktt}{{k_T}}
\newcommand{\kt}{{\mathbf{k}_T}}
\newcommand{\ktone}{{\mathbf{k}_{T1}}}
\newcommand{\kttwo}{{\mathbf{k}_{T2}}}
\newcommand{\bt}{{\mathbf{b}_T}}
\newcommand{\xt}{{\mathbf{x}_T}}
\newcommand{\yt}{{\mathbf{y}_T}}
\newcommand{\rt}{{\mathbf{r}_T}}
\newcommand{\rtp}{{\mathbf{r}'_T}}

\newcommand{\qs}{Q_{\textnormal{s}}}

\newcommand{\fig}{Fig.~}

\newcommand{\eq}{Eq.~}

\usepackage{amssymb}
\usepackage{amsmath}

\title{Small-x, Diffraction and Vector Mesons}
\ShortTitle{Small-x, Diffraction and Vector Mesons}

\author{\speaker{T. Lappi}\\
Department of Physics, 
 P.O. Box 35, 40014 University of Jyv\"askyl\"a, Finland
\\
Helsinki Institute of Physics, P.O. Box 64, 00014 University of Helsinki,
Finland \\
E-mail: \email{tuomas.v.v.lappi@jyu.fi}}

\abstract{This talk discusses recent progress in some
topics relevant for deep inelastic
scattering at small~$x$. We discuss first
differences and similarities between conventional collinear factorization 
and the dipole picture of deep inelastic scattering. Many of the recent
 theoretical advances at small~$x$ are related to taking calculations 
in the nonlinear saturation regime to next-to-leading order accuracy 
in the QCD coupling. On the experimental 
side significant recent progress has been made in exclusive and 
diffractive processes, 
in particular in ultraperipheral nucleus-nucleus collisions.}

\FullConference{XXIII International Workshop on Deep-Inelastic 
Scattering\\
                 27 April - May 1 2015\\
                 Dallas, Texas}

\begin{document}

\section{Introduction}

Our understanding of DIS at small~$x$ is evolving rapidly due to theoretical advances, 
current and anticipated experimental progress and exploitation of the
complementarity between different
processes. It is impossible to try to give an exhaustive review here, and therefore
this paper will focus on three core topics. We will first discuss the differences 
and similarities between the collinear factorization picture used by the largest
part of the community represented at the DIS 2015 conference  and the dipole picture
that is very convenient for understanding QCD in the small $x$ saturation limit. 
We will then move to a biased selection of the topics that are discussed
also elsewhere 
in these proceedings. Two of the important advances in small-$x$ theory recently 
have been the significant progress in pushing cross section calculations to the 
next-to-leading order accuracy in the QCD coupling also in the nonlinear 
saturation regime, 
and opening up new connections between small~$x$ physics and spin physics. On the
experimental side, in addition to final HERA analyses and preparatory work for a
new generation of high energy DIS experiments, there are many interesting LHC results
on ultraperipheral (i.e. photon-proton or photon-nucleus) 
collisions that are very relevant for this context. The discussion 
here will stay in the regime of weak coupling physics, unfortunately leaving aside 
recent progress in soft diffraction and central exclusive production.

\section{Different views of small-$x$ physics}

\subsection{Dipole picture and collinear factorization}

\begin{wrapfigure}{R}{0.3\textwidth}
\includegraphics[width=0.3\textwidth]{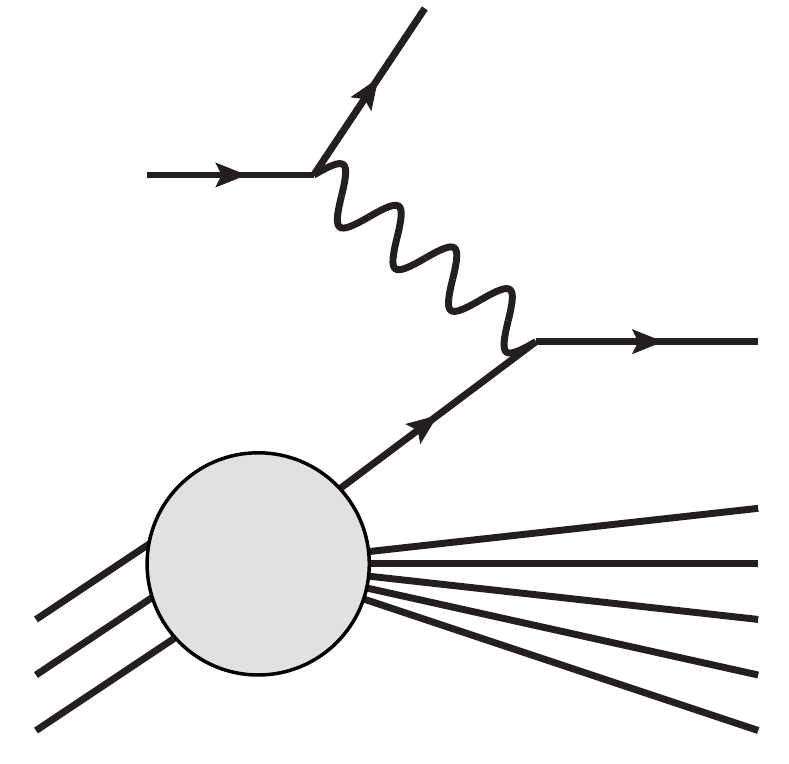}
\caption{DIS process in the IMF: a virtual photon interacts with a pre-existing
quark in the target.}
\label{fig:dislocoll}
\end{wrapfigure}
In QCD one cannot calculate cross sections exactly. Instead, one arranges 
the calculations in a perturbative series assuming a small value for the 
coupling constant. When the scattering process studied involves QCD bound states, as 
is the case for DIS, one needs to separate the part of the process that 
can be described using weak coupling from the nonperturbative hadronic physics. The
latter must then be parametrized by some functions of appropriate kinematical 
variables which must  ultimately be obtained from experimental data or, in 
principle, from lattice calculations. A priori, there is no unique way to decide 
what are the best degrees of freedom to describe the nonperturbative part of
the process. Different scattering processes or different kinematical regimes might
be more optimally described using a different language. In order for the theory 
to have predictive power, the nonperturbative description should be as universal as 
possible. This means that the same degrees of freedom can be used in descriptions
of different processes, e.g. measured in one process to calculate a genuine prediction
for the cross section of another. It is also helpful, if not strictly speaking necessary, 
for the description to be related to a simple physical picture of the nonperturbative
QCD bound state. 
A cross section is naturally a Lorentz invariant quantity and as such independent
of the frame in which one decides to view the process. The microscopic degrees of
freedom describing a QCD bound state can, however, be very different  in 
different Lorentz frames. These different ways of viewing the nonperturbative
structure of the bound state can  be naturally adapted to very 
different schemes of organizing perturbative QCD calculations.
This is indeed the case for the Infinite Momentum Frame (IMF) 
and the Target Rest Frame (TRF) (more 
appropriately the dipole frame) that we will describe in the following. Both are
used to organize QCD calculations into perturbatively calculable parts, a nonperturbative 
input and weak coupling renormalization group equations describing the dependence
of the nonperturbative input on some kinematical variable. Both 
have their advantages and disadvantages and are applicable for partially overlapping
sets of scattering processes and kinematical regimes.

\begin{wrapfigure}{R}{0.4\textwidth}
\includegraphics[width=0.4\textwidth]{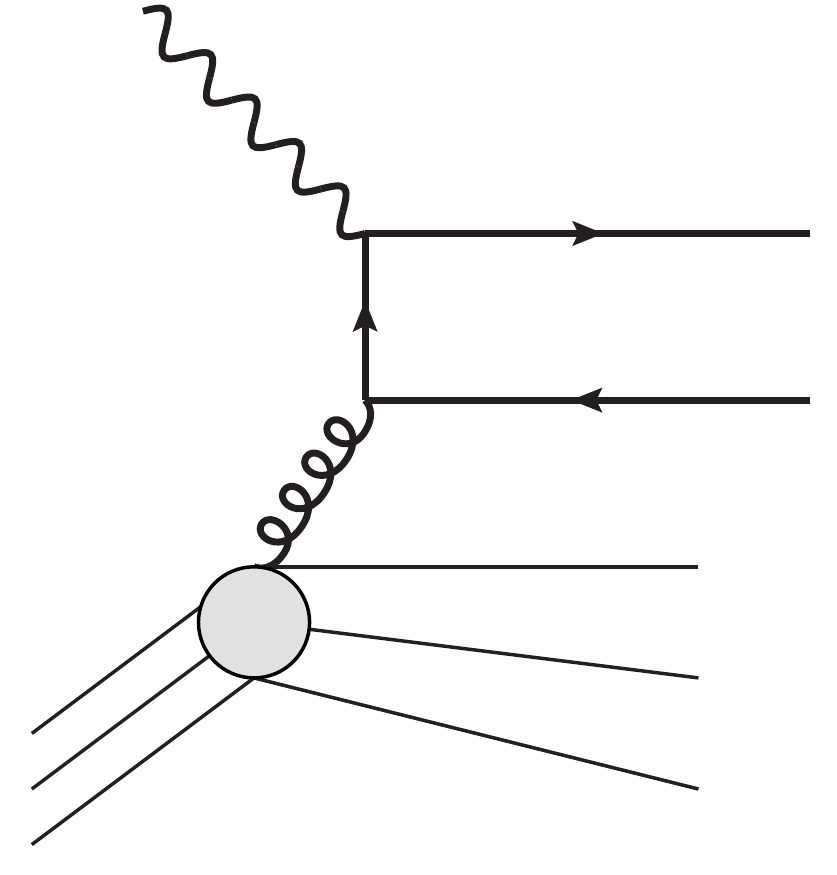}
\caption{Higher order DIS process in the IMF: a virtual photon interacts with 
gluon via a quark-antiquark pair.}
\label{fig:disnlocoll}
\end{wrapfigure}
The usual perturbative QCD description of a DIS process is based on the
collinear factorization formalism. The most natural physical picture in this
case is the one in the IMF, where the proton moves at a very high energy. The relevant 
degrees of freedom in the proton are then quasi-free partons, quarks and gluons. The
partons are collinear, i.e. they share a fraction $x=k^+/ P^+$ of the longitudinal
momentum (more precisely $+$-momentum) of the proton, but have a very small 
transverse momentum $\ktt$. According to a simple uncertainty argument the (light cone)
lifetime $\Delta x^+ \sim 1/k^- = 2k^+/\ktt^2$  of these fluctuations is 
long compared to
the resolution scale of the virtual photon, $\Delta x^+ \sim 1/q^- \sim k^+/Q^2$.
Thus the process can be viewed as a virtual photon instantaneously measuring the 
partonic content of the proton; to leading order just the quark distribution, 
see \fig\ref{fig:dislocoll}.  Using light cone quantization, one can relate 
the cross section to the Fock state decomposition of the proton.

\begin{wrapfigure}{R}{0.4\textwidth}
\includegraphics[width=0.4\textwidth]{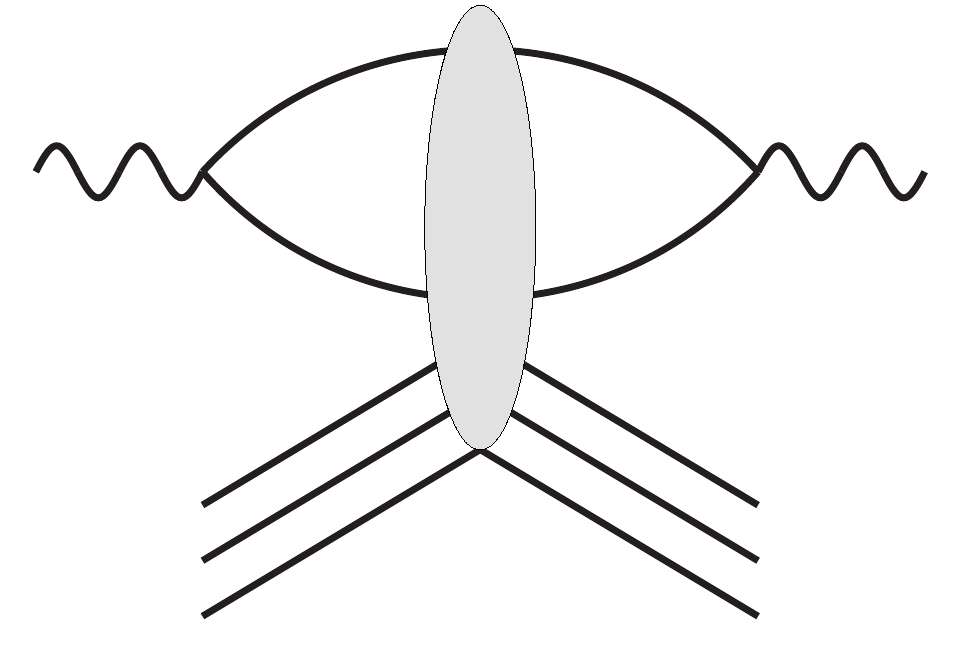}
\caption{Leading order DIS process in the dipole picture: the virtual photon splits into
a quark-antiquark pair, which interacts with the target.}
\label{fig:dislodip}
\end{wrapfigure}
To leading order in perturbation theory, the virtual photon only measures the 
quark number density in the target proton, see \fig\ref{fig:dislocoll}.
QCD dynamics is probed in the collinear picture when one goes beyond leading order,
e.g. in the process shown in \fig\ref{fig:disnlocoll}.
The degrees of freedom used to parametrize the nonperturbative physics of the proton
are number densities of quarks and gluons, i.e. parton distribution functions (PDF's). 
By extracting the leading large transverse 
momentum $Q^2$ behavior of the higher order diagrams such as 
\fig\ref{fig:disnlocoll}, one can derive the DGLAP renormalization equations,
which describe the dependence of these PDF's on the external virtuality $Q^2$. 
The kinematical regime of validity of
the decription for sufficiently inclusive processes 
is very broad; the longitudinal momentum fraction $x$ can be arbitrary
as long as the momentum scale $Q^2$ is large enough.

Now let us look at the same process in a different frame where the target proton moves 
slowly, but the virtual photon has a large momentum $q^+$. Now, according to exactly the
same kinematical argument as above, quantum fluctuations in the virtual photon 
exist on a much longer timescale than the interaction. They are then instantaneously struck by the color
fields in the target proton. During this short interaction time the structure of the probe 
$\gamma^*$ is frozen. A natural method to understand this process is to 
light cone quantize the 
virtual photon.
Since it is a clean and 
controlled object and not a complicated QCD bound state, 
the light cone quantization  part of the calculation 
is purely perturbative. The nonperturbative object describing the target,
on the other hand, is its scattering amplitude with the probe.
The microscopical description of the target naturally associated with this
picture is as a classical strong color field. Since the probe has
a high energy, its interaction with this field can be calculated using the
eikonal approximation.

\begin{wrapfigure}{r}{0.3\textwidth}
\includegraphics[width=0.3\textwidth]{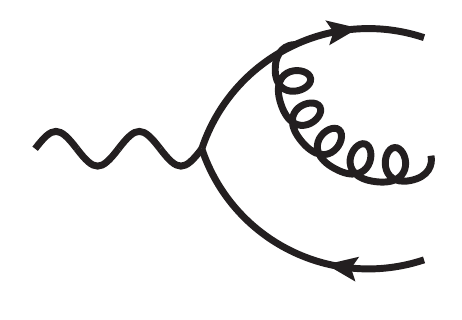}
\caption{Next to leading order DIS process in the dipole picture: 
the virtual photon state also has a quark-antiquark-gluon component.}
\label{fig:disnlodip}
\end{wrapfigure}
 The lowest nontrivial component in the $\gamma^*$ state that can  interact
with a target color field is a quark-antiquark dipole, \fig\ref{fig:dislodip}. 
As in the collinear picture, QCD dynamics comes in when one moves to higher order 
in perturbation theory. In this case the perturbative expansion is the one describing the 
virtual photon, and the NLO correction involves the inclusion of an additional gluon in the 
virtual photon state, as depicted in \fig\ref{fig:disnlodip}.
 At least in the large $\nc$ limit, or more generally using a
more explicit model for the scattering, e.g. the classical field picture, the scattering
amplitude of the $q\bar{q}g$ state can be related to that of the original dipole. The
integration over the phase space of the additional gluon yields a correction to the
cross section of the original dipole that is enhanced by a large logarithm of the
energy $W^2$. 
These logarithms can then be resummed into a renormalization 
group equation: BK~\cite{Balitsky:1995ub,Kovchegov:1999yj} or JIMWLK~\cite{Jalilian-Marian:1997xn,Jalilian-Marian:1997jx,Jalilian-Marian:1997gr,%
Jalilian-Marian:1997dw,JalilianMarian:1998cb,Iancu:2000hn,%
Iancu:2001md,Ferreiro:2001qy,Iancu:2001ad,Weigert:2000gi,Mueller:2001uk} that describes the dependence of the scattering amplitude on the energy.
The procedure is completely analogous to the physics of DGLAP, with $\ln 1/x$ 
playing the role of $\ln Q^2$ as the large logarithm resummed and the scattering amplitude
the role of the PDF's as the function describing the target. The downside of this 
picture is that it is only applicable at small $x$, with $Q^2$ not too large but 
large enough for the coupling to be weak. But to compensate for this there are several 
advantages. Since one is working directly with the imaginary part of the forward elastic 
scattering amplitude, i.e. the total dipole-target cross section, implementing unitarity
constraints is straighforward and transparent. The same amplitude is 
also probed in elastic processes, i.e. diffractive DIS,
which enhances the predictive power of the framework. This should be
contrasted with  collinear perturbation theory, where 
diffractive PDF's are usually fit to data independently from the inclusive ones. 
Also, since one is not counting gluons in the 
target but treating them coherently as a color field, the formalism is much more naturally 
suited to describing the nonlinear physics of gluon saturation.

\subsection{Gluon saturation} 

The physical picture of renormalization group evolution in the collinear
 framework
is one of a cascade, where partons split by gluon emission and pair creation. 
As the virtuality $Q^2$ or energy $\sim 1/x$ increases, 
there is phase space for more of these splittings, and the number of partons grows.
At some point in the cascade it is possible that the phase space
density of gluons is large enough for also mergings of gluons to become important.
We can parametrically estimate when this happens using the following simple argument~\cite{Mueller:1985wy}.
We can assume, by the uncertainty principle, that one gluon accupies an area
$\sim 1/Q^2$ in the tranverse plane. 
The number of these gluons per unit rapidity in a proton of
area $\pi \rp^2$ is
given by the gluon distribution $xG(x,Q^2)$. Thus the number of gluons in a cell of 
size $1/Q^2$ is given by the gluon density times the area of the cell $\sim 
(xG(x,Q^2)/\pi \rp^2) \times (1/Q^2)$. For gluon mergings to be important
they must overlap enough to pay the price
of the coupling $\as$ in the probability for merging. This leads to the estimate
that the gluon mergings become important when
$xG(x,Q^2)/(Q^2 \pi \rp^2) \gtrsim 1/\as$. The $x$-dependent value of the 
characteristic transverse  momentum scale at which this happens is known as the 
\emph{saturation scale} $\qs$; it is the solution of the equation
\begin{wrapfigure}{r}{0pt}
\includegraphics[height=3cm]{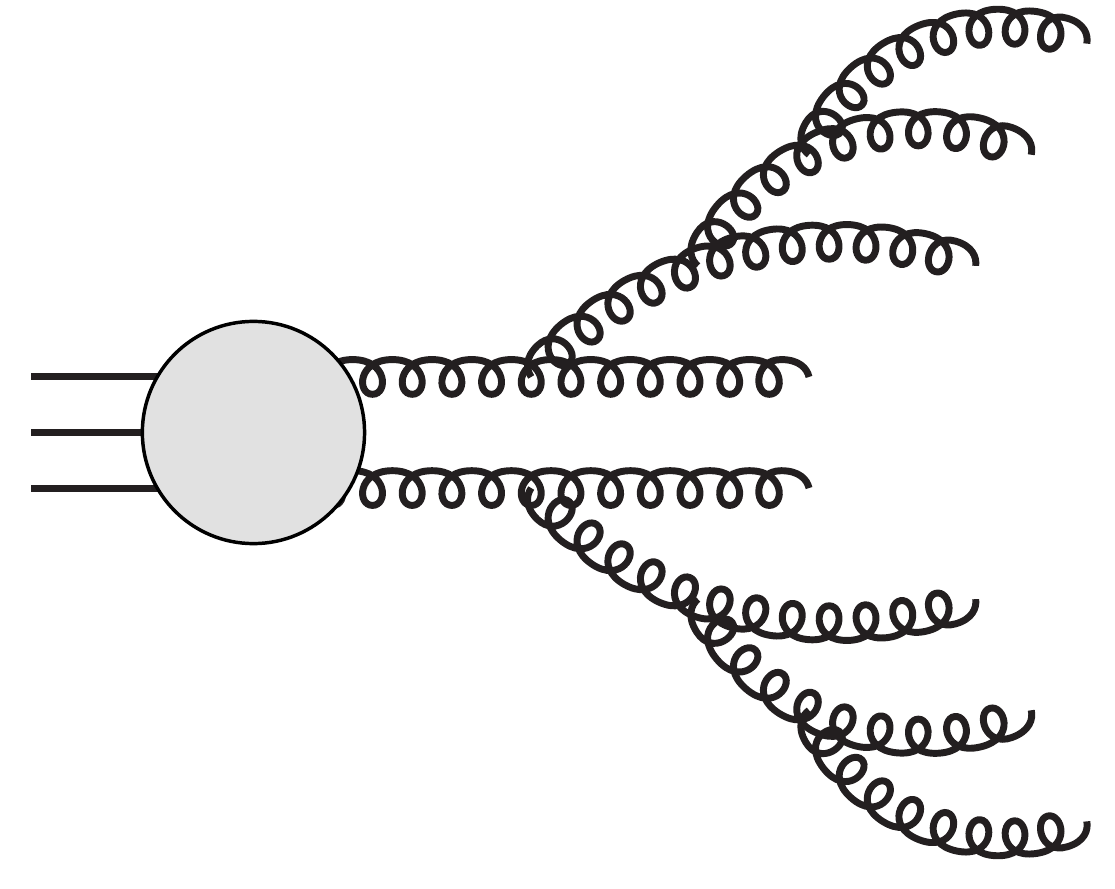}
\includegraphics[height=3cm]{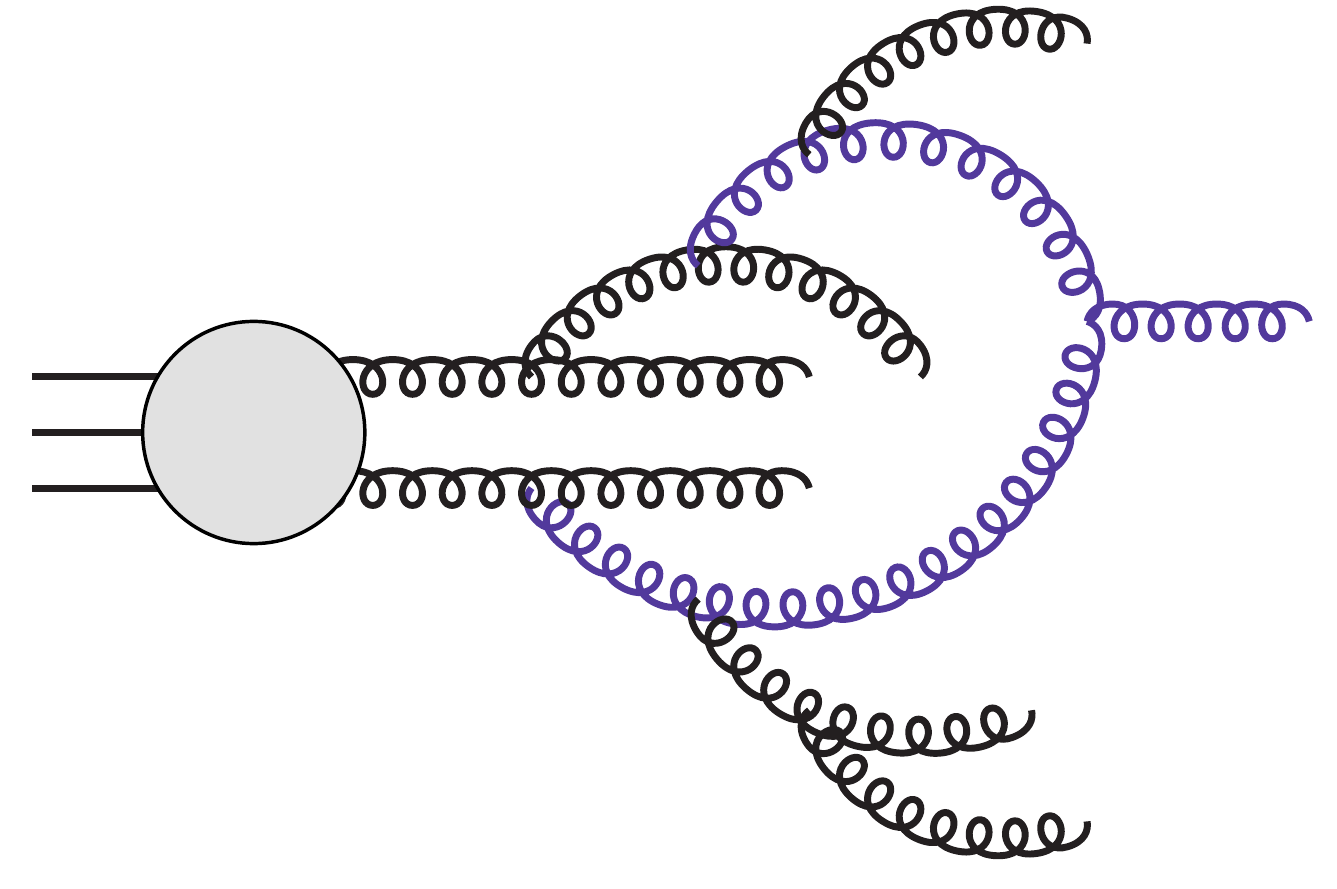}
\caption{Linear and nonlinear cascade of gluons}
\label{fig:cascade}
\end{wrapfigure}
\begin{equation}
\pi R_p^2  \qs^2 \sim \as xG(x,\qs^2) .
\label{eq:satcond}
\end{equation}
This equation is naturally an order of magnitude estimate, and the exact 
numerical value  of a characteristic momentum scale depends 
on the precise definition. We will argue in the following that it is more natural 
to give such a definition in the dipole picture.
Diagrammatically this transition from a cascade of pure splittings 
to one which also 
includes mergings could be illustrated by something like \fig\ref{fig:cascade}.
In practice it is difficult to work quantitatively with gluon merging diagrams
of the kind sketched in \fig\ref{fig:cascade}
(however, for an interesting attempt to introduce saturation into a Monte Carlo 
cascade see~\cite{Avsar:2006jy}). Thus the IMF collinear picture is  not
very convenient for understanding saturation quantitatively. 
It is also not
clear whether the chosen degree of freedom, a number density of gluons, is
conceptionally meaningful in a regime of important nonlinear interactions between 
them.

In the dipole picture the same phenomenon of parton saturation appears in a very 
different way. The target proton is described by a dimensionless (imaginary part of the 
forward elastic) 
scattering amplitude $\ampli$, which is related to the total dipole-target 
cross section by the optical theorem
$\sigma = 2 \int \ud^2  \bt\ampli$. The amplitude naturally varies between 
no interaction $\ampli=0$ and total absorption, i.e. the black disk limit $\ampli=1$ 
(in principle unitarity allows values up to $\ampli=2$).
We are working in the eikonal approximation where the size of the dipole $r\sim 1/Q$
 stays 
fixed during the interaction. Since a  dipole of size $r=0$ is a color neutral object 
and should not interact in QCD, we  know that $\ampli(r=0)=0$. For small enough
dipoles the scattering should be weak and 
dominated by the perturbatively 
leading process of two-gluon exchange. Indeed one has 
\begin{equation}
\sigma \sim 2 \pi R_p^2 \mathcal{N}(r) \sim  \as r^2 xG 
\left(x,Q^2\sim 1/r^2\right)
\quad \textnormal{ for } \quad 
r\to 0 .
\label{eq:smallrdipxs}
\end{equation}
Now we can immediately see that the perturbative two-gluon exchange approximation 
cannot remain valid for arbitrarily large dipoles, i.e. small $Q^2$. Some
mechanism involving multiple gluon ecxhanges must begin to be important 
when $\ampli \sim \as r_\textnormal{s}^2 xG(x,Q^2\sim 1/r_\textnormal{s}^2) / (\pi \rp^2) \sim 1$. With the identification $\qs = 1/r_\textnormal{s}$
this is exactly the same saturation condition that we obtained previously in 
\eq\nr{eq:satcond}. The difference is that now the phenomenon does not appear
to follow from nonlinear gluon dynamics. Instead it is required by unitarity 
and must, and easily can, be built  into the formalism used. 
This is the case in the CGC framework, and in particular 
for the BK and JIMWLK equations.
In practice this is done by assuming that the
target is described by a strong color field, with which
the quark and antiquark in the dipole interact 
eikonally by picking up an SU($\nc$) Wilson line
\begin{equation}
V = \mathbb{P} \exp\left\{ -ig\int\!\! \ud x^+ A^-\right\}.
\label{eq:defV}
\end{equation}
The dipole amplitude then turns out to be a correlation function 
of two of these Wilson lines
\begin{equation}
\ampli(\rt = \xt-\yt) = 1-\frac{1}{\nc} \bigg< \tr V^\dag(\xt)V(\yt) \bigg>,
\label{eq:CGCdipxs}
\end{equation}
automatically satisfying the requirements
$\ampli(r=0)=0$ and $\ampli(r\to \infty)=1$ (since at large separations the
Wilson lines are uncorrelated and average separately to zero for strong fields).

\section{NLO theory}

\subsection{Evolution equations}

\begin{wrapfigure}{R}{0pt}
\includegraphics[width=3cm]{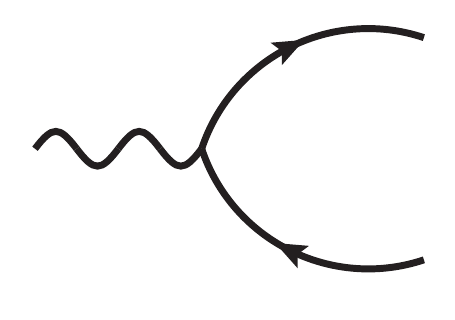}
\includegraphics[width=3cm]{gammatoqqbarg1}
\includegraphics[width=3cm]{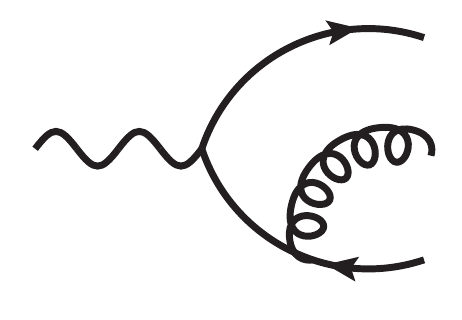}
\begin{tikzpicture}[overlay,scale=1.5]
\node[anchor=west]at (-4.85,0.7) {$\rt$};
\node[anchor=west]at (-2.6,1) {$\rtp$};
\node[anchor=west]at (-0.1,0.5) {$\rt-\rtp$};
\draw[line width=1pt,<->] (-4.8,0.35) -> (-4.8,1.15);
\draw[line width=1pt,<->] (-2.55,0.75) -> (-2.55,1.2);
\draw[line width=1pt,<->] (-0.05,0.3) -> (-0.05,0.7);
\end{tikzpicture}
\rule{3em}{0pt}
\caption{Diagrams needed to derive the BK equation.}
\label{fig:bk}
\end{wrapfigure}
As discussed above, by including one additional gluon to the dipole (see \fig\ref{fig:bk}), relating
the $q\bar{q}$ and $q\bar{q}g$ cross sections to each other using the large $\nc$ 
mean field limit and picking up the leading logarithm of $x$ from the
integration over the phase space of the gluon one can derive the 
leading order BK equation for the $y=\ln 1/x$-dependence of the scattering
amplitude $\mathcal{N}$:
\begin{equation}
\partial_y \mathcal{N}(\rt)
= 
 \frac{\as \nc}{2\pi^2}  
\int \ud^2 \rtp \frac{\rt^2}{\rtp^2\left(\rtp-\rt\right)^2}
\left[
 \mathcal{N}\left(\rtp\right)
+
 \mathcal{N}\left(\rt-\rtp\right)
- 
\mathcal{N}\left(\rtp\right)
 \mathcal{N}\left(\rt-\rtp\right)
- \mathcal{N}\left(\rt\right)
\right] .
\end{equation}
In the linear limit (dropping the $\ampli^2$-term) this reduces to the LO BFKL equation.
The BFKL equation has been known to NLO accuracy for some time \cite{Fadin:1996nw,Fadin:1998py,Ciafaloni:1998gs}. At the
NLO level one starts also picking up large logarithms of transverse momenta, which
were subsequently  resummed in Mellin 
space~\cite{Salam:1998tj,Ciafaloni:1999yw,Altarelli:1999vw,Ciafaloni:2003rd} 
in order to get a meaningful result. 
The NLO version of the nonlinear BK equation~\cite{Balitsky:2008zza} (see also 
the equation~\cite{Balitsky:2014mca} for the baryon operator) has also been known for
several years. A full numerical solution of this equation, was, however, 
performed only very recently~\cite{Lappi:2015fma} (see, however, previous 
numerical  studies~\cite{Avsar:2011ds}). As could be expected, it
 suffers from the same problems with large
transverse logarithms as NLO BFKL. Resumming them in the case of the nonlinear 
equation requires, however, different methods that are only now being 
developed~\cite{Iancu:2015joa,Iancu:2015vea}. At the time of the DIS 2015 conference,
the state of the art for phenomenological applications is still the leading order
BK equation, supplemented with running coupling corrections.

The BK equation uses a mean-field approximation (replacing 
$\left<\hat{D}\hat{D} \right>\to \left<\hat{D}\right>\left<\hat{D} \right>$,
where $\hat{D}=\frac{1}{\nc}\tr V^\dag V$ is the dipole operator)
to close the equation. Without this approximation one obtains a more general
infinite hierarchy
of equations that couples expectation values of different products of Wilson lines. This 
hierarchy can be equivalently written as a renormalization group equation for the 
\emph{probability distribution} of Wilson lines, and is known as the JIMWLK equation.
While the NLO JIMWLK equation cannot directly be read off from the NLO BK equation,
the loop integrals required in deriving the two are mostly the same. This has 
very recently made it possible to derive the NLO JIMWLK 
equation~\cite{Balitsky:2013fea,Kovner:2013ona}, relying heavily on these earlier 
results.

\subsection{Forward particle production in pA}

\begin{figure}
\centerline{\includegraphics[height=3.5cm]{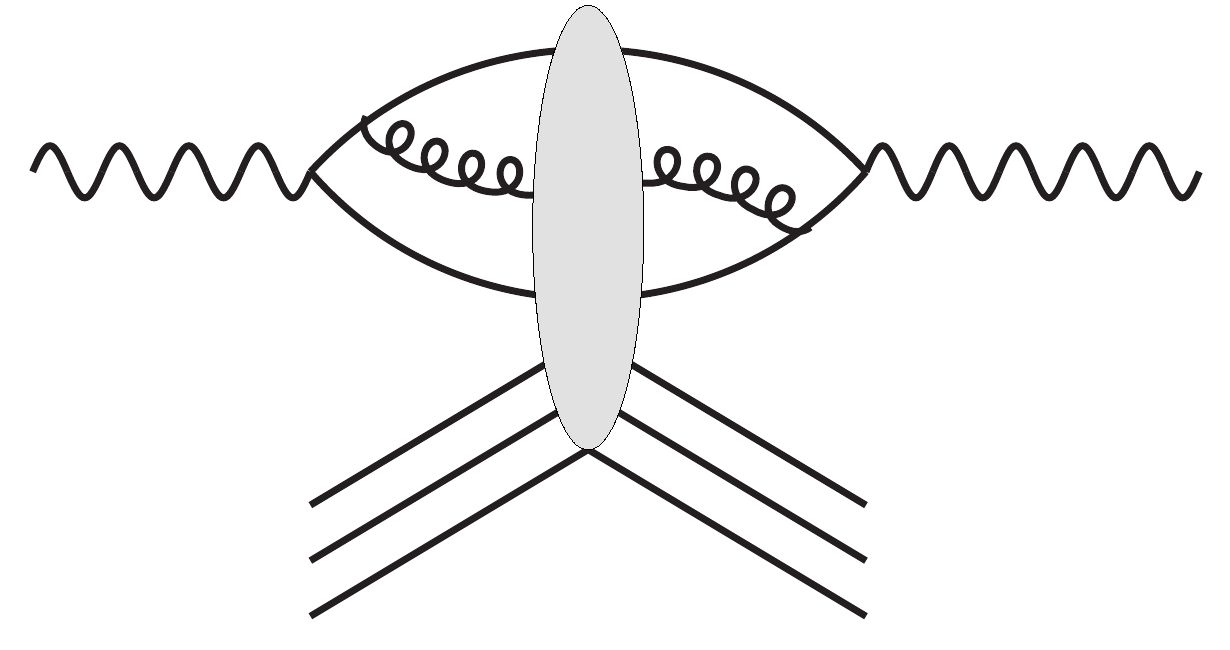}
\rule{3em}{0pt}\includegraphics[height=3.5cm]{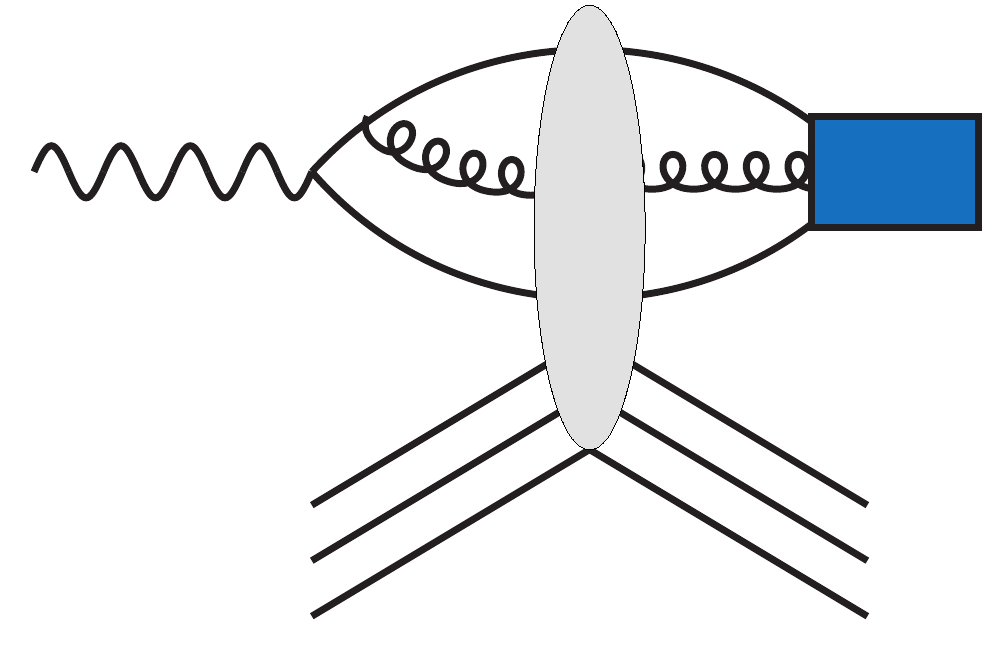}}
\caption{Left: A contribution to the NLO impact factor for DIS.
Right: A contribution to NLO exclusive vector meson production. }
\label{fig:disifnlo}
\label{fig:vmnlo}
\end{figure}

For a full NLO calculation of a physical cross section the evolution 
equations by themselves are not enough. One must also calculate the parts of the process 
without a large resummed logarithm to NLO accuracy. For the case of DIS this 
means the full calculation of diagrams such as the one in \fig\ref{fig:disifnlo},
where the leading high energy logarithm gives the LO evolution equation and the 
remainder is a part of the NLO ''impact factor''. This has been
done in two separate papers~\cite{Balitsky:2010ze,Beuf:2011xd} using a slightly 
different formalism. The consistency of these two results with each other
is yet to be confirmed, and neither of them has yet been applied to practical 
phenomenology. A similar calculation, involving the modeling of the $q\bar{q}g$ state
in a vector meson wavefunction (see \fig\ref{fig:vmnlo}) would be required 
for exclusive vector meson production.

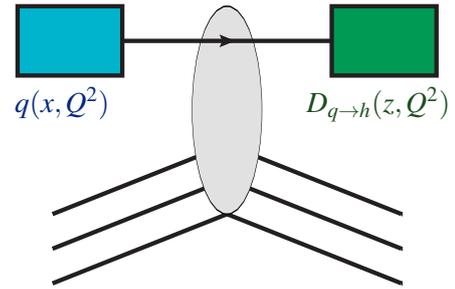
\begin{wrapfigure}{R}{0pt}
\includegraphics[width=6cm]{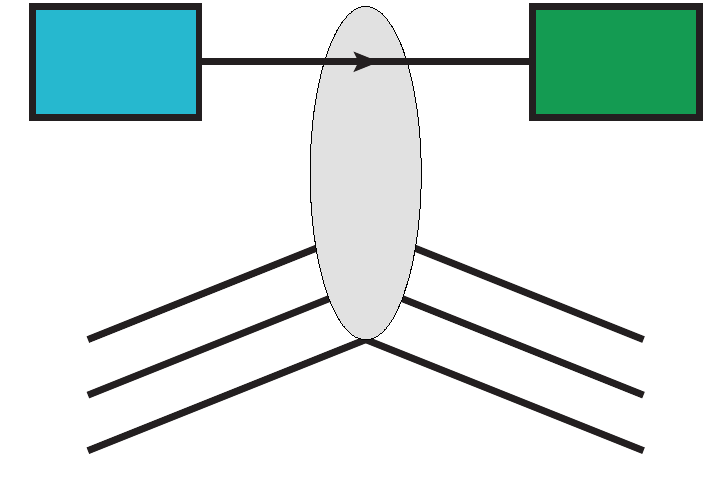} 
\begin{tikzpicture}[overlay,scale=2]
\node[anchor=north west]at (-3,1.5) {\textcolor{omasin}{$q(x,Q^2)$}};
\node[anchor=north east]at (0,1.5) {\textcolor{omavihrea}{$D_{q\to h}(z,Q^2)$}};
\end{tikzpicture}
\caption{Forward particle production at leading order.}
\label{fig:palo}
\end{wrapfigure}
While progress in NLO DIS phenomenology has been slower, there has been much 
activity recently on calculating  particle production 
in forward proton-nucleus collisions to NLO.
Here the physical picture (see \fig\ref{fig:palo})
starts from a high-$x$ quark or gluon  from the proton, described using a conventional collinear PDF. The quark
passes through the strong color field of the target nucleus,
acquiring transverse momentum from the intrinsic~$\ktt$ of the small-$x$ gluons in
the target and finally fragments into hadrons. The fragmentation can be described by 
conventional fragmentation functions, but the interaction with the target 
is given
by the same Wilson line as the quark and antiquark in DIS, \eq\nr{eq:defV}. When
the eikonal quark-target scattering amplitude is squared, one obtains a
cross section which is essentially given by just the Fourier-transform of the (DIS) 
dipole cross section \nr{eq:CGCdipxs}.
\begin{figure}[t]
\centerline{\includegraphics[height=3.5cm]{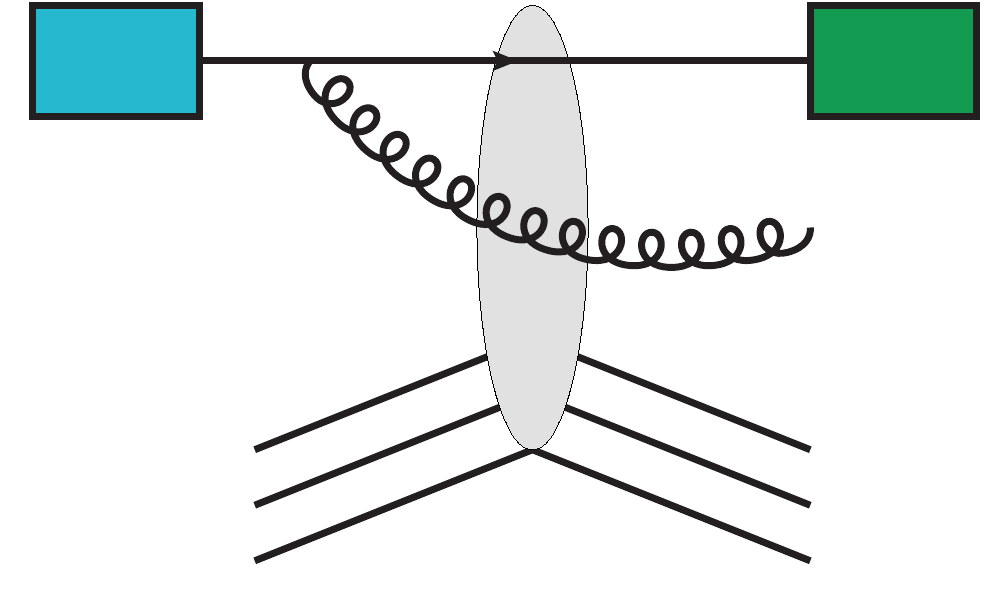}
\rule{3em}{0pt}\includegraphics[height=3.5cm]{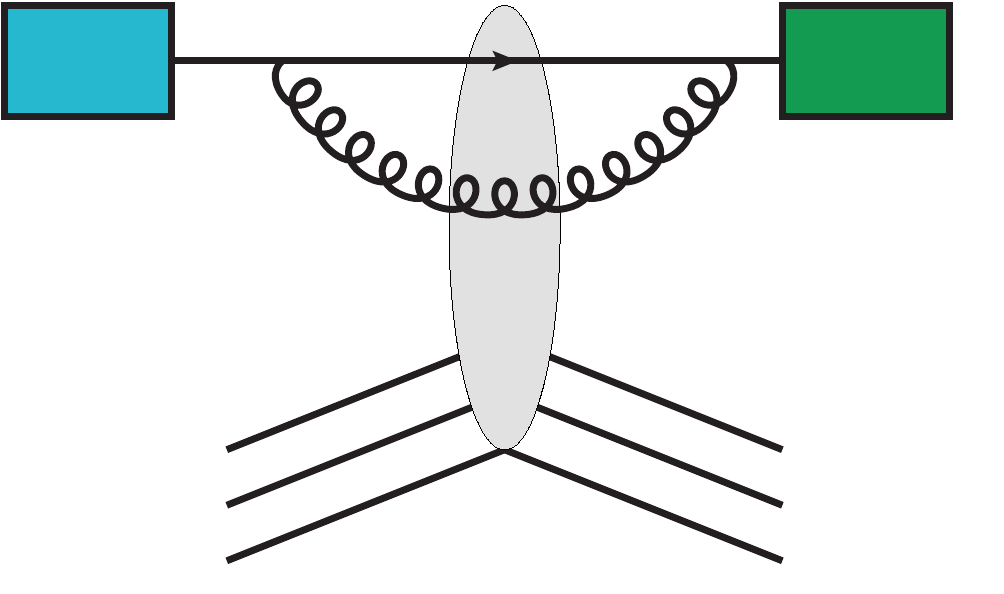}}
\caption{Real (left) and virtual (right) NLO corrections to quark production 
in forward proton-nucleus collisions.}
\label{fig:panlo}
\end{figure}

\rule{0pt}{1pt}

\begin{wrapfigure}{R}{0pt}
\includegraphics[width=0.5\textwidth]{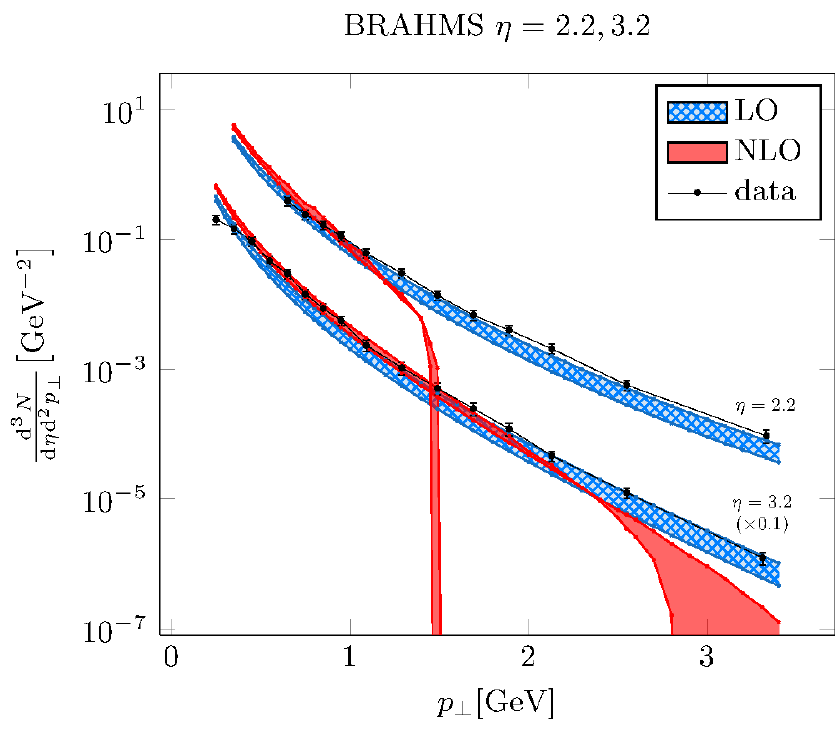} 
\caption{LO and NLO single inclusive cross sections from \cite{Stasto:2013cha}
compared to BRAHMS data.}
\label{fig:pastasto}
\end{wrapfigure} 
At NLO accuracy one needs to consider the one loop corrections, e.g. the ones 
shown in 
\fig\ref{fig:panlo}. Some years back Chirilli 
et. al.~\cite{Chirilli:2011km,Chirilli:2012jd} demonstrated that, as expected,
the appropriate logs
of $x$ and $Q^2$ in these contributions can be factorized into the BK 
evolution of the target amplitude and the DGLAP evolution of the probe PDF and
the fragmentation function. Unfortunately attempts to apply these results led to 
very large NLO corrections that, at high $\ptt$, made the total cross section 
negative~\cite{Stasto:2013cha} (see \fig\ref{fig:pastasto}). Several authors have tried to understand this 
problem~\cite{Stasto:2014sea,Kang:2014lha,Altinoluk:2014eka,Watanabe:2015tja} and propose solutions.
While some progress has indeed been made on the matter, it seems fair to say
that the question of NLO particle production is still unsettled, and still an area
of active study.

\subsection{Transverse momentum and spin}

The conventional operator definition for the transverse momentum 
dependent (TMD) gluon distribution is
\begin{equation}
 xG(x,\kt) = \int   \frac{\ud x^- \ud^2 \xt}{(2\pi)^3 P^+}
e^{ixP^+x^- - i \kt\cdot\xt}
\langle P | F^{+i}(x^-,\xt) \mathcal{L} F^{+i}(0) |P\rangle .
\end{equation}
Since the operator is nonlocal, the two insertions
of the field strength tensor $F^{i+}$ have to be connected by a
gauge link $\mathcal{L}$, which follows a path that can be different 
for different processes. With a more general Lorentz index 
structure one can also write down similar expressions for
spin dependent distributions. A major advance in the field in recent years
has been to understand~\cite{Dominguez:2011wm,Kovchegov:2013cva,Balitsky:2015qba,Kotko:2015ura,Kovchegov:2015zha}
 how these different TMD distributions can, at small $x$, 
be expressed in terms of the Wilson lines~\nr{eq:defV}. In the conventional
perturbation theory framework the different distributions must be separately
fit to separate experimental data. Being able to express them all in terms
of the same Wilson line greatly enhances the universality and predictive
power of the theory. In the CGC framework one can, for example, use only
$F_2$ structure function data to fit the initial conditions for 
the BK evolution of the distribution of Wilson lines. These can then be used 
to calculate various TMD and spin distributions
without any additional experimental data. 

One example of such a TMD distribution is
the linearly polarized gluon distribution~\cite{Metz:2011wb}.
Experimentally it is probed in dijet production in DIS,
where the dependence  of the cross sections on the azimuthal 
angle between the total and relative  
transverse momenta of the two jets (i.e. the angle between 
$\ktone+\kttwo$ and $\ktone-\kttwo$, where $\ktone$ and $\kttwo$ 
are the transverse momenta of the jets). 
Measurements of  multiparticle azimuthal correlations
are a subject of much interest also in heavy ion collisions,
where they are created in abundance by collective effects in 
the plasma. Indeed, starting from a coordinate space asymmetric
initial distribution of matter as in a noncentral heavy ion collision,
hydrodynamical flow generates strong anisotropies in momentum 
space due to the anisotropy of pressure gradients. The similar
experimental signature of these very different physical effects, 
collective flow in QCD matter and polarization and correlation effects 
that are present already in DIS, makes the interpretation of
multiparticle correlations measurements in proton-proton and 
proton-nucleus collisions very challenging.

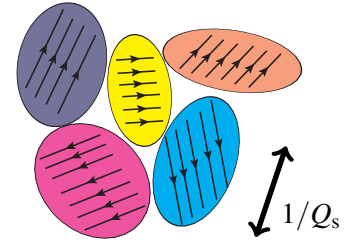
\begin{wrapfigure}{R}{0pt}
\includegraphics[width=4cm]{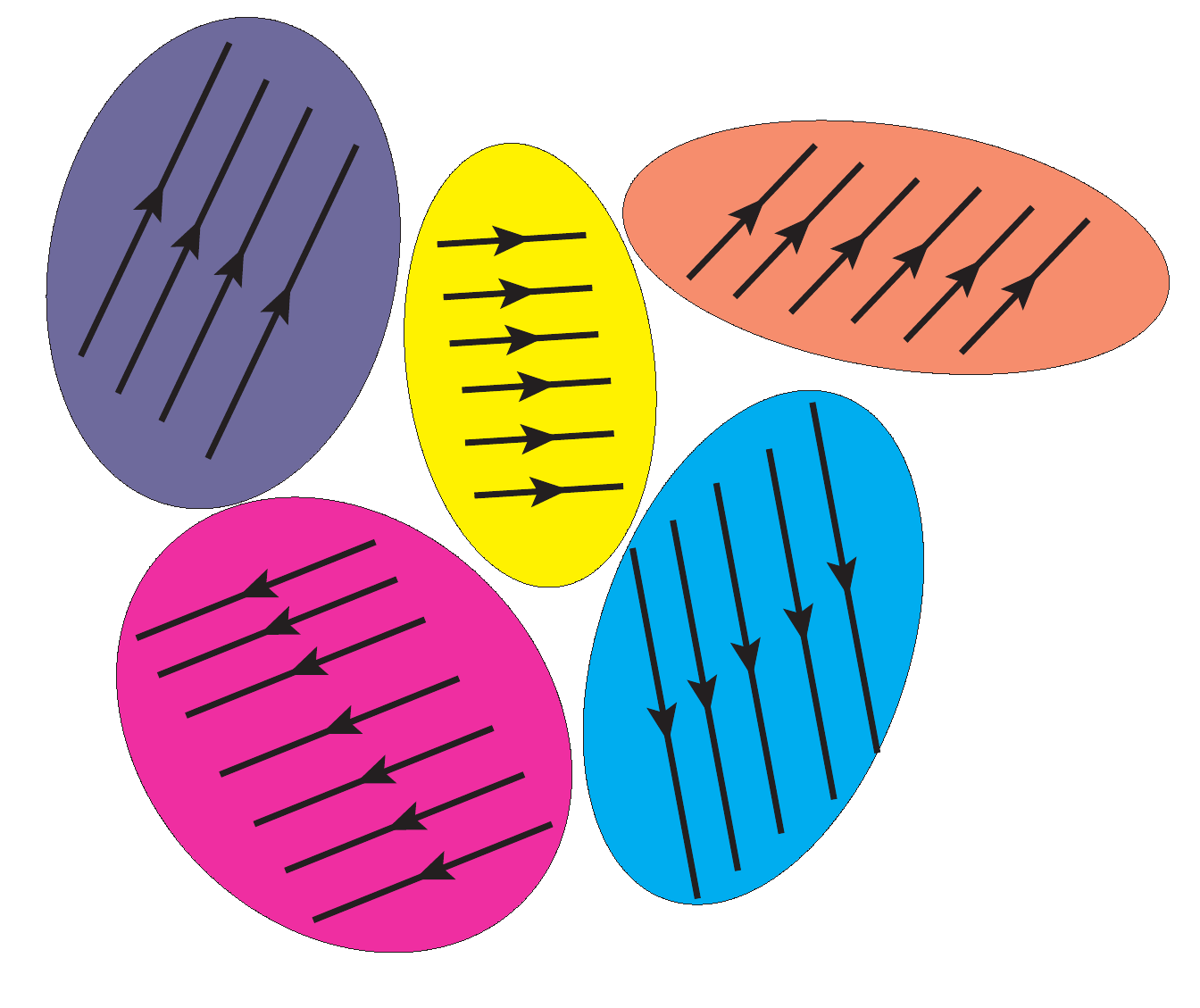} 
\begin{tikzpicture}[overlay]
\node[anchor=north west]at (-0.6,0.8) {$1/\qs$};
\draw[line width=2pt,<->](-0.8cm,0.2cm) -- (-0.4cm,1.4cm);
\end{tikzpicture}
\caption{Color fields seen by dilute probe.}
\label{fig:domains}
\end{wrapfigure}
In the CGC picture, the mechanism for generating azimuthal correlations
whithout collective final state effects
has a very intuitive physical interpretation, which 
is easiest to describe in the case of a dilute object (dipole, 
forward proton) colliding with the dense color field~\cite{Kovner:2010xk}.
The picture of particle production in this case is that of
the particles from the probe passing through target that consists
of domains of transverse color electric field (\fig\ref{fig:domains}).
The size of these domains in coordinate space is $\sim 1/\qs$, which 
corresponds to the target consisting of gluons with $\ktt \sim \qs$.
Particles from the probe that pass through the same domain in the
target and that have the same color experience the same color 
electric field and are deflected in the same direction. This naturally
leads to an angular correlation between these particles, and
has been put forward~\cite{Dumitru:2008wn,Dumitru:2010iy,Dumitru:2014dra,Lappi:2015vha} as an at least partial explanation 
for many such correlation phenomena seen in proton-proton and proton-nucleus 
collisions.

\section{Exclusive processes}

\begin{wrapfigure}{r}{0pt}
\includegraphics[height=2.5cm]{newdipole}
\begin{tikzpicture}[overlay]
\draw[omapun,line width=2pt] (-1.9,1.6) ellipse (0.4 and 1);
\end{tikzpicture}
\includegraphics[height=2.5cm]{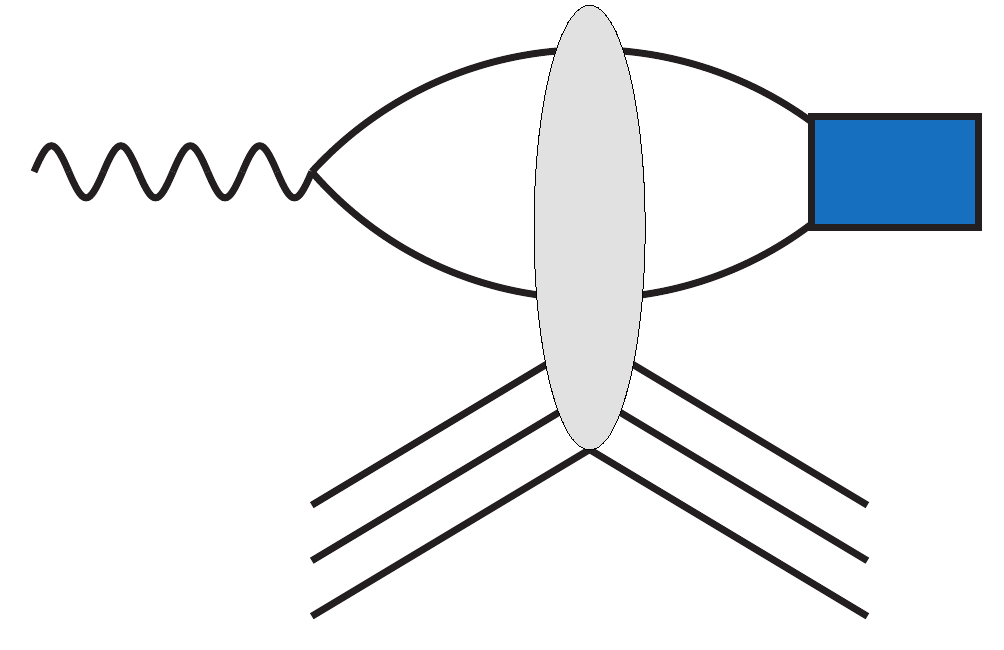}
\begin{tikzpicture}[overlay]
\draw[omapun,line width=2pt] (-1.7,1.6) ellipse (0.4 and 1);
\draw[omasin,dashed,line width=2pt] (-0.5,1.85) ellipse (0.6 and 0.4);
\node[coordinate] (lcpic) at (-0.2,1.5){cc};
\end{tikzpicture}
\caption{Total DIS cross section via the optical theorem (left) and 
the amplitude for exclusive vector meson production (right). The circled
blob is the common amplitude, while the  box surrounded by the dashed circle
is the vector meson light cone wavefunction.}
\label{fig:vm}
\end{wrapfigure}

Let us finally move to a  topic where recent progress has been more
experimentally driven, namely exclusive processes and in particular
vector meson production. One of the strengths of the dipole picture
is that the cross section is given by the square of the same
elastic dipole-target scattering amplitude that determines the total 
cross section. This enables a direct relation between inclusive
diffraction and the total cross 
section~\cite{Nikolaev:1991et,Kowalski:2003hm,Kowalski:2006hc}. 
The additional ingredient needed for the 
description of  exclusive vector meson production
 is the transition matrix element
from the dipole to the vector meson bound state. This leads one to 
focus attention on 
heavy quark mesons, whose bound state properties can to some
degree be understood perturbatively. In light cone quantization
this information is expressed in terms of the so called vector
meson (light cone) ``wavefunction'' (see \fig\ref{fig:vm}).

Exclusive processes  can provide important constraints for
models of small~$x$ gluons. However, due to the additional
complication of the vector meson bound states, systematical
studies are needed  as a function 
of $x$, $Q^2$ (see e.g.~\cite{Kovalchuk:2014hea}), vector meson
species and nuclear mass number. 
While waiting for an Electron-Ion Collider to 
explore the full parameter space, the nuclear mass number dependence 
can be addressed in the photoproduction limit $Q^2=0$ by ultraperipheral
nucleus-nucleus collisions (UPC's) at the LHC and at RHIC.
The idea with UPC is to use the high electric charge of the fully 
ionized nucleus as a source of real photons. By triggering on events
with no hadronic activity (besides the  vector meson) one
restricts the impact parameter to be beyond the range
of the strong interaction; hence the term ultraperipheral. 
Using the equivalent flux of Weizs\"{a}cker-Williams
photons from the ion one can relate the measured cross section
to the photon-nucleus (or photon-proton) cross section.

Experimental studies of exclusive vector meson production 
are often portrayed as measurements of the nuclear gluon distribution.
This interpretation is based on the formula that relates the exclusive
vector meson cross section to the target gluon distribution as
\begin{equation}
\left. \frac{\ud \sigma^{\gamma^*H \to V H}}{\ud t} \right|_{t=0}= 
\frac{16 \pi^3 \as^2 \Gamma_{ee}}{3 \alpha_{\textnormal{em}} M_V^5}
\left[xg(x,Q^2=M_V^2)\right]^2.
\end{equation}
This equation is in fact~\cite{Brodsky:1994kf} the small $r$
limit of the dipole model calculation, where the dipole cross
section is related to the gluon distribution as in \eq\nr{eq:smallrdipxs}
and the wavefunction is reduced to its behavior around the origin, 
where it can be related to the electromagnetic
decay width of the meson $\Gamma_{ee}$. Some of the theory calculations
are indeed based on this formula, usually supplemented with phenomenological
corrections for the normalization.  In the CGC picture, however, it is 
natural to go beyond the small-$r$ limit and include the full integral
over different size dipoles.

\begin{figure}[tb]
\centerline{\includegraphics[height=4cm]{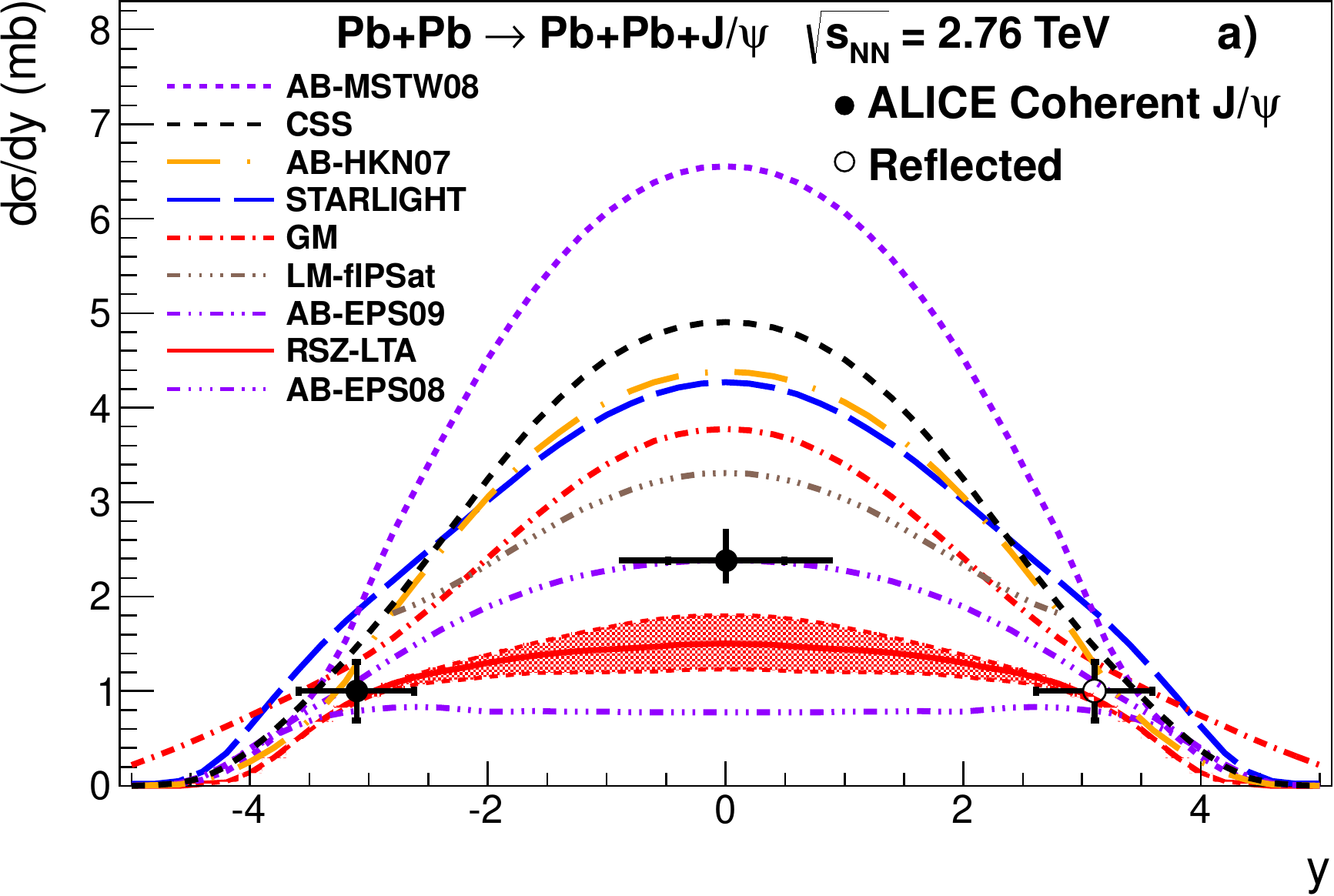}
\includegraphics[height=4cm]{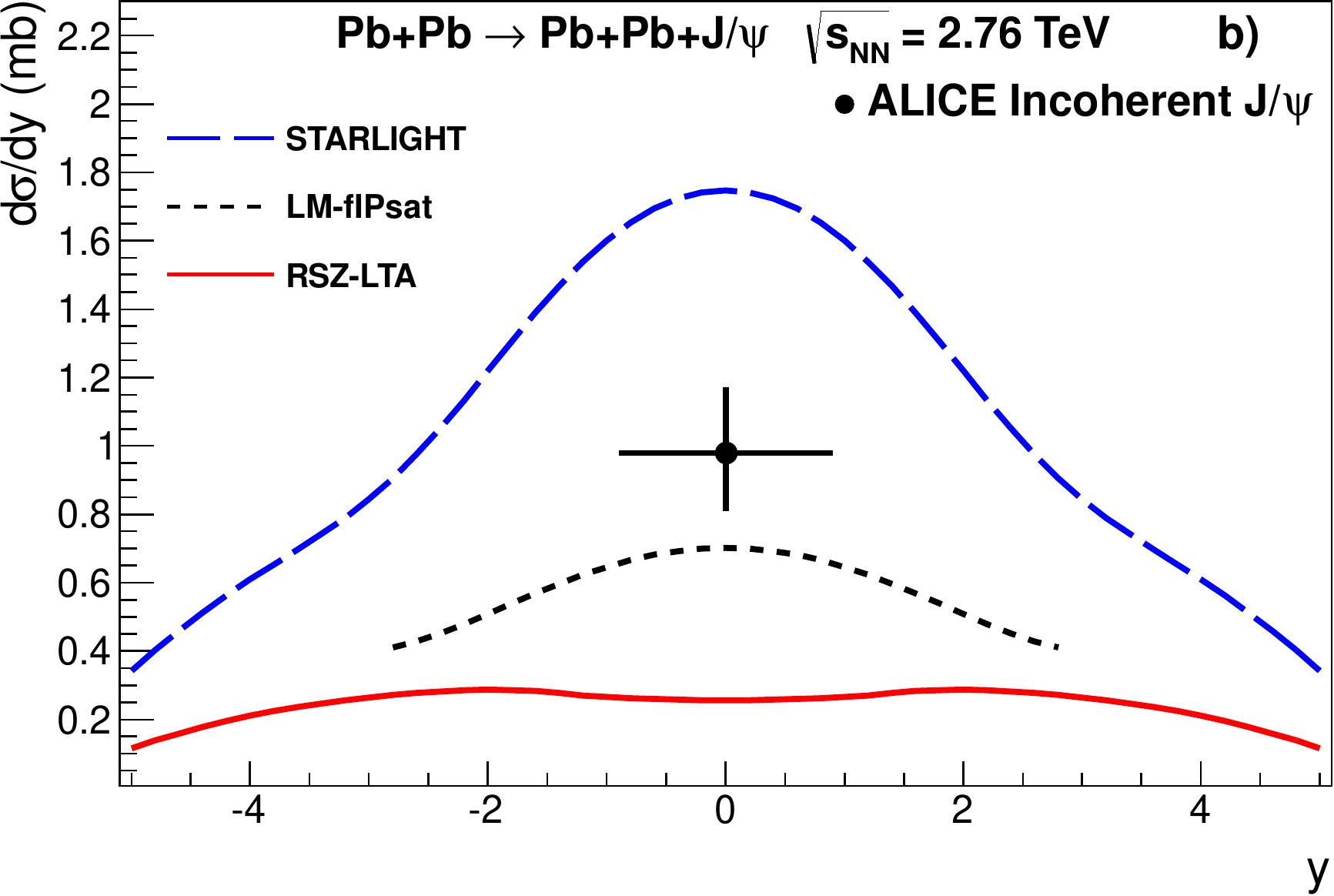}
}
\caption{ALICE results for coherent $J/\Psi$ production compared
to theoretical models, from \cite{Abbas:2013oua}}
\label{fig:aliceupc}
\end{figure}

In the case of nuclear targets one distinguishes two
classes of events: \emph{coherent}, where the nucleus
stays intact in its ground state, and \emph{incoherent}, where
it breaks up, but only into color neutral protons and neutrons, preserving
the rapidity gap characteristic of diffractive events.
Coherent events dominate for small momentum transfers to the target
$|t|\sim 1/\ra^2$, where $\ra$ is the nuclear radius. For 
$|t|\sim 1/\rp^2$, which is the typical momentum transfer
for exclusive $ep$ events, the incoherent process dominates. 
Even here, however, the cross section is very much affected
by nuclear effects, being suppressed by a factor 
$\sim 3$~\cite{Lappi:2010dd} compered to $A$ times the nucleon
cross section. This suppression 
can be understood as arising from a ``survival probability'', where
 the (virtual) photon, after interacting elastically with one nucleon, 
must  \emph{not} interact inelastically with spectator nucleons
to preserve the rapidity gap.
The ALICE collaboration distinguishes between these two classes largely 
based on the transverse momentum of the vector meson which, for $Q^2=0$, 
is the same as the recoil momentum of the nucleus.
The coherent cross section is related to the average gluon density
in the nucleus, while the incoherent one directly measures
the fluctuations~\cite{Lappi:2010dd}. This interpretation is very
implicit in the SARTRE event generator~\cite{Toll:2012mb}.
Thus the incoherent data provides an important
 constraint on the fluctuating nuclear geometry that 
has become an important aspect of interpretations of heavy ion collisions,
and deserves to be better addressed by the
theory community.

\begin{wrapfigure}{R}{0pt}
\includegraphics[width=0.5\textwidth]{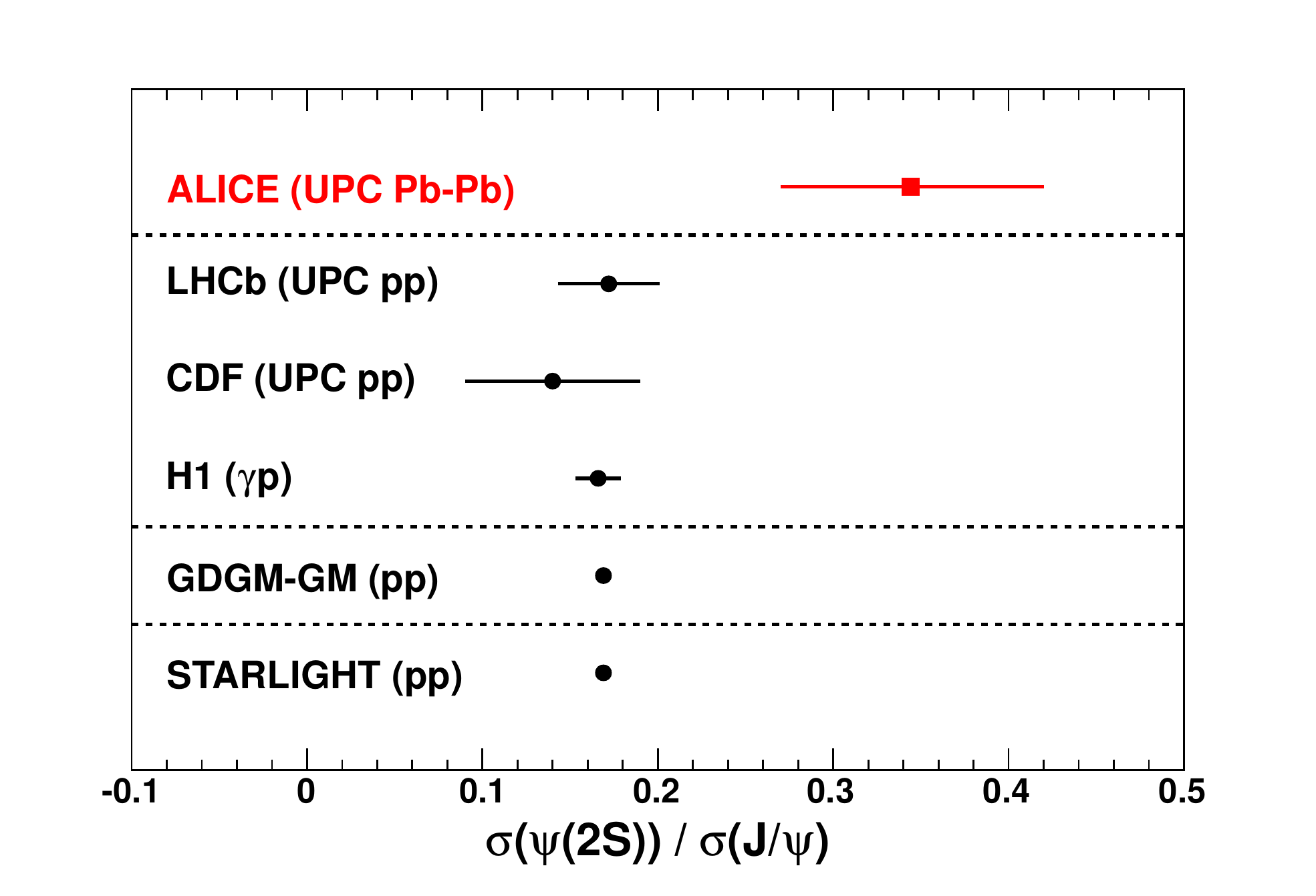}
\caption{
Ratio of $J/\Psi$ to $\Psi(2S)$ photoproduction cross sections
from $\gamma$-proton collisions compared to the
ALICE result from $\gamma$-nucleus collisions~\cite{Adam:2015sia}.
}
\label{fig:aliceratio}
\end{wrapfigure}
Figure~\ref{fig:aliceupc} shows first results on $J/\Psi$ production 
from ALICE~\cite{Abbas:2013oua,Abelev:2012ba}
compared to theory calculations. Results have also been published by
LHCb~\cite{Aaij:2014iea} ($\gamma$-proton collisions) and 
CMS~\cite{CMS:2014ies} ($\gamma$-nucleus collisions).
A particularly interesting result from ALICE is on the  $\Psi(2S)$ to $J/\Psi$ 
cross section ratio (see \fig\ref{fig:aliceratio}), which
is measured to be approximately twice as large as in $\gamma p$ collisions. 
In the dipole picture
the cross section of the $2S$ state is suppressed with respect to the
$1S$ state due to the ``node effect'', i.e. a cancellation between 
different sign parts in the excited state meson wavefunction. 
Since the increased saturation scale diminishes the relative
importance of the large-$r$ cross section, this node effect
cancellation is naturally weaker for nuclei than for protons. 
In practice, however, it is difficult to see how the effect 
could become as large as indicated by the ALICE data.
With many of the measurements
still statistics limited, we can expect many more such interesting 
results from the future LHC runs.

\section{Conclusions}

In conclusion, we have here discussed some recent focus areas in small-$x$ 
physics, concentrating in particular on developments in the 
nonlinear gluon saturation regime. On the theory side, there is a systematical
effort to push calculations to NLO accuracy. In spite of the recent progress
there is still work to do to understand the interplay of small~$x$ and 
large~$Q^2$ logarithms at the NLO level. Another related area
have been connections between the saturation formalism and that of
TMD gluon distributions. Also here, the understanding of 
the formal theory side has improved significantly, but  there is still much 
to do to turn these ideas into precision phenomenology. We have also discussed
recent experimental progress, in particular with exclusive photoproduction 
of vector mesons at the LHC.

\paragraph{Acknowledgements}
 This work has been supported by the Academy of Finland, projects 
267321 and 273464. The author thanks K.~J.~Eskola for comments on the manuscript.

\bibliography{spires}

\providecommand{\href}[2]{#2}\begingroup\raggedright\begin{thebibliography}{10}

\bibitem{Balitsky:1995ub}
I.~Balitsky, {\it Operator expansion for high-energy scattering},
  \href{http://dx.doi.org/10.1016/0550-3213(95)00638-9}{{\em Nucl. Phys.} {\bf
  B463} (1996) 99} [\href{http://arXiv.org/abs/hep-ph/9509348}{{\tt
  arXiv:hep-ph/9509348}}].

\bibitem{Kovchegov:1999yj}
Y.~V. Kovchegov, {\it Small-x {F2} structure function of a nucleus including
  multiple pomeron exchanges},
  \href{http://dx.doi.org/10.1103/PhysRevD.60.034008}{{\em Phys. Rev.} {\bf
  D60} (1999) 034008} [\href{http://arXiv.org/abs/hep-ph/9901281}{{\tt
  arXiv:hep-ph/9901281}}].

\bibitem{Jalilian-Marian:1997xn}
J.~Jalilian-Marian, A.~Kovner, L.~D. McLerran and H.~Weigert, {\it The
  intrinsic glue distribution at very small x},
  \href{http://dx.doi.org/10.1103/PhysRevD.55.5414}{{\em Phys. Rev.} {\bf D55}
  (1997) 5414} [\href{http://arXiv.org/abs/hep-ph/9606337}{{\tt
  arXiv:hep-ph/9606337 [hep-ph]}}].

\bibitem{Jalilian-Marian:1997jx}
J.~Jalilian-Marian, A.~Kovner, A.~Leonidov and H.~Weigert, {\it The {BFKL}
  equation from the {Wilson} renormalization group},
  \href{http://dx.doi.org/10.1016/S0550-3213(97)00440-9}{{\em Nucl. Phys.} {\bf
  B504} (1997) 415} [\href{http://arXiv.org/abs/hep-ph/9701284}{{\tt
  arXiv:hep-ph/9701284}}].

\bibitem{Jalilian-Marian:1997gr}
J.~Jalilian-Marian, A.~Kovner, A.~Leonidov and H.~Weigert, {\it The {Wilson}
  renormalization group for low x physics: Towards the high density regime},
  \href{http://dx.doi.org/10.1103/PhysRevD.59.014014}{{\em Phys. Rev.} {\bf
  D59} (1999) 014014} [\href{http://arXiv.org/abs/hep-ph/9706377}{{\tt
  arXiv:hep-ph/9706377}}].

\bibitem{Jalilian-Marian:1997dw}
J.~Jalilian-Marian, A.~Kovner and H.~Weigert, {\it The {Wilson} renormalization
  group for low x physics: Gluon evolution at finite parton density},
  \href{http://dx.doi.org/10.1103/PhysRevD.59.014015}{{\em Phys. Rev.} {\bf
  D59} (1999) 014015} [\href{http://arXiv.org/abs/hep-ph/9709432}{{\tt
  arXiv:hep-ph/9709432}}].

\bibitem{JalilianMarian:1998cb}
J.~Jalilian-Marian, A.~Kovner, A.~Leonidov and H.~Weigert, {\it Unitarization
  of gluon distribution in the doubly logarithmic regime at high density},
  \href{http://dx.doi.org/10.1103/PhysRevD.59.034007}{{\em Phys. Rev.} {\bf
  D59} (1999) 034007} [\href{http://arXiv.org/abs/hep-ph/9807462}{{\tt
  arXiv:hep-ph/9807462}}].

\bibitem{Iancu:2000hn}
E.~Iancu, A.~Leonidov and L.~D. McLerran, {\it Nonlinear gluon evolution in the
  color glass condensate. {I}},
  \href{http://dx.doi.org/10.1016/S0375-9474(01)00642-X}{{\em Nucl. Phys.} {\bf
  A692} (2001) 583} [\href{http://arXiv.org/abs/hep-ph/0011241}{{\tt
  arXiv:hep-ph/0011241}}].

\bibitem{Iancu:2001md}
E.~Iancu and L.~D. McLerran, {\it Saturation and universality in {QCD} at small
  x},  \href{http://dx.doi.org/10.1016/S0370-2693(01)00526-3}{{\em Phys. Lett.}
  {\bf B510} (2001) 145} [\href{http://arXiv.org/abs/hep-ph/0103032}{{\tt
  arXiv:hep-ph/0103032}}].

\bibitem{Ferreiro:2001qy}
E.~Ferreiro, E.~Iancu, A.~Leonidov and L.~McLerran, {\it Nonlinear gluon
  evolution in the color glass condensate. {II}},
  \href{http://dx.doi.org/10.1016/S0375-9474(01)01329-X}{{\em Nucl. Phys.} {\bf
  A703} (2002) 489} [\href{http://arXiv.org/abs/hep-ph/0109115}{{\tt
  arXiv:hep-ph/0109115}}].

\bibitem{Iancu:2001ad}
E.~Iancu, A.~Leonidov and L.~D. McLerran, {\it The renormalization group
  equation for the color glass condensate},
  \href{http://dx.doi.org/10.1016/S0370-2693(01)00524-X}{{\em Phys. Lett.} {\bf
  B510} (2001) 133} [\href{http://arXiv.org/abs/hep-ph/0102009}{{\tt
  arXiv:hep-ph/0102009}}].

\bibitem{Weigert:2000gi}
H.~Weigert, {\it Unitarity at small {Bjorken} x},
  \href{http://dx.doi.org/10.1016/S0375-9474(01)01668-2}{{\em Nucl. Phys.} {\bf
  A703} (2002) 823} [\href{http://arXiv.org/abs/hep-ph/0004044}{{\tt
  arXiv:hep-ph/0004044 [hep-ph]}}].

\bibitem{Mueller:2001uk}
A.~H. Mueller, {\it A simple derivation of the {JIMWLK} equation},
  \href{http://dx.doi.org/10.1016/S0370-2693(01)01343-0}{{\em Phys. Lett.} {\bf
  B523} (2001) 243} [\href{http://arXiv.org/abs/hep-ph/0110169}{{\tt
  arXiv:hep-ph/0110169}}].

\bibitem{Mueller:1985wy}
A.~H. Mueller and J.-w. Qiu, {\it Gluon recombination and shadowing at small
  values of x},  \href{http://dx.doi.org/10.1016/0550-3213(86)90164-1}{{\em
  Nucl. Phys.} {\bf B268} (1986) 427}.

\bibitem{Avsar:2006jy}
E.~Avsar, G.~Gustafson and L.~Lonnblad, {\it Small-x dipole evolution beyond
  the {large-$N_\textnormal{c}$} limit},
  \href{http://dx.doi.org/10.1088/1126-6708/2007/01/012}{{\em JHEP} {\bf 01}
  (2007) 012} [\href{http://arXiv.org/abs/hep-ph/0610157}{{\tt
  arXiv:hep-ph/0610157 [hep-ph]}}].

\bibitem{Fadin:1996nw}
V.~S. Fadin and L.~Lipatov, {\it Next-to-leading corrections to the {BFKL}
  equation from the gluon and quark production},
  \href{http://dx.doi.org/10.1016/0550-3213(96)00334-3}{{\em Nucl. Phys.} {\bf
  B477} (1996) 767} [\href{http://arXiv.org/abs/hep-ph/9602287}{{\tt
  arXiv:hep-ph/9602287 [hep-ph]}}].

\bibitem{Fadin:1998py}
V.~S. Fadin and L.~Lipatov, {\it {BFKL} pomeron in the next-to-leading
  approximation},  \href{http://dx.doi.org/10.1016/S0370-2693(98)00473-0}{{\em
  Phys. Lett.} {\bf B429} (1998) 127}
  [\href{http://arXiv.org/abs/hep-ph/9802290}{{\tt arXiv:hep-ph/9802290
  [hep-ph]}}].

\bibitem{Ciafaloni:1998gs}
M.~Ciafaloni and G.~Camici, {\it Energy scale(s) and next-to-leading {BFKL}
  equation},  \href{http://dx.doi.org/10.1016/S0370-2693(98)00551-6}{{\em Phys.
  Lett.} {\bf B430} (1998) 349}
  [\href{http://arXiv.org/abs/hep-ph/9803389}{{\tt arXiv:hep-ph/9803389
  [hep-ph]}}].

\bibitem{Salam:1998tj}
G.~Salam, {\it A resummation of large subleading corrections at small x},
  \href{http://dx.doi.org/10.1088/1126-6708/1998/07/019}{{\em JHEP} {\bf 9807}
  (1998) 019} [\href{http://arXiv.org/abs/hep-ph/9806482}{{\tt
  arXiv:hep-ph/9806482 [hep-ph]}}].

\bibitem{Ciafaloni:1999yw}
M.~Ciafaloni, D.~Colferai and G.~Salam, {\it Renormalization group improved
  small x equation},  \href{http://dx.doi.org/10.1103/PhysRevD.60.114036}{{\em
  Phys. Rev.} {\bf D60} (1999) 114036}
  [\href{http://arXiv.org/abs/hep-ph/9905566}{{\tt arXiv:hep-ph/9905566
  [hep-ph]}}].

\bibitem{Altarelli:1999vw}
G.~Altarelli, R.~D. Ball and S.~Forte, {\it Resummation of singlet parton
  evolution at small x},
  \href{http://dx.doi.org/10.1016/S0550-3213(00)00032-8}{{\em Nucl. Phys.} {\bf
  B575} (2000) 313} [\href{http://arXiv.org/abs/hep-ph/9911273}{{\tt
  arXiv:hep-ph/9911273 [hep-ph]}}].

\bibitem{Ciafaloni:2003rd}
M.~Ciafaloni, D.~Colferai, G.~Salam and A.~Stasto, {\it Renormalization group
  improved small x {Green's} function},
  \href{http://dx.doi.org/10.1103/PhysRevD.68.114003}{{\em Phys. Rev.} {\bf
  D68} (2003) 114003} [\href{http://arXiv.org/abs/hep-ph/0307188}{{\tt
  arXiv:hep-ph/0307188 [hep-ph]}}].

\bibitem{Balitsky:2008zza}
I.~Balitsky and G.~A. Chirilli, {\it Next-to-leading order evolution of color
  dipoles},  \href{http://dx.doi.org/10.1103/PhysRevD.77.014019}{{\em Phys.
  Rev.} {\bf D77} (2008) 014019} [\href{http://arXiv.org/abs/0710.4330}{{\tt
  arXiv:0710.4330 [hep-ph]}}].

\bibitem{Balitsky:2014mca}
I.~Balitsky and A.~V. Grabovsky, {\it {NLO} evolution of 3-quark {Wilson} loop
  operator},  \href{http://dx.doi.org/10.1007/JHEP01(2015)009}{{\em JHEP} {\bf
  01} (2015) 009} [\href{http://arXiv.org/abs/1405.0443}{{\tt arXiv:1405.0443
  [hep-ph]}}].

\bibitem{Lappi:2015fma}
T.~Lappi and H.~M{\"a}ntysaari, {\it Direct numerical solution of the
  coordinate space {Balitsky-Kovchegov} equation at next to leading order},
  \href{http://dx.doi.org/10.1103/PhysRevD.91.074016}{{\em Phys. Rev.} {\bf
  D91} (2015) 074016} [\href{http://arXiv.org/abs/1502.02400}{{\tt
  arXiv:1502.02400 [hep-ph]}}].

\bibitem{Avsar:2011ds}
E.~Avsar, A.~Stasto, D.~Triantafyllopoulos and D.~Zaslavsky, {\it
  Next-to-leading and resummed {BFKL} evolution with saturation boundary},
  \href{http://dx.doi.org/10.1007/JHEP10(2011)138}{{\em JHEP} {\bf 1110} (2011)
  138} [\href{http://arXiv.org/abs/1107.1252}{{\tt arXiv:1107.1252 [hep-ph]}}].

\bibitem{Iancu:2015joa}
E.~Iancu, J.~D. Madrigal, A.~H. Mueller, G.~Soyez and D.~N. Triantafyllopoulos,
  {\it Collinearly-improved {BK} evolution meets the {HERA} data},
  \href{http://arXiv.org/abs/1507.03651}{{\tt arXiv:1507.03651 [hep-ph]}}.

\bibitem{Iancu:2015vea}
E.~Iancu, J.~Madrigal, A.~Mueller, G.~Soyez and D.~Triantafyllopoulos, {\it
  Resumming double logarithms in the {QCD} evolution of color dipoles},
  \href{http://arXiv.org/abs/1502.05642}{{\tt arXiv:1502.05642 [hep-ph]}}.

\bibitem{Balitsky:2013fea}
I.~Balitsky and G.~A. Chirilli, {\it Rapidity evolution of {Wilson} lines at
  the next-to-leading order},
  \href{http://dx.doi.org/10.1103/PhysRevD.88.111501}{{\em Phys. Rev.} {\bf
  D88} (2013) 111501} [\href{http://arXiv.org/abs/1309.7644}{{\tt
  arXiv:1309.7644 [hep-ph]}}].

\bibitem{Kovner:2013ona}
A.~Kovner, M.~Lublinsky and Y.~Mulian, {\it {Jalilian-Marian}, {Iancu},
  {McLerran}, {Weigert}, {Leonidov}, {Kovner} evolution at next to leading
  order},  \href{http://dx.doi.org/10.1103/PhysRevD.89.061704}{{\em Phys. Rev.}
  {\bf D89} (2014) 061704} [\href{http://arXiv.org/abs/1310.0378}{{\tt
  arXiv:1310.0378 [hep-ph]}}].

\bibitem{Balitsky:2010ze}
I.~Balitsky and G.~A. Chirilli, {\it Photon impact factor in the
  next-to-leading order},
  \href{http://dx.doi.org/10.1103/PhysRevD.83.031502}{{\em Phys. Rev.} {\bf
  D83} (2011) 031502} [\href{http://arXiv.org/abs/1009.4729}{{\tt
  arXiv:1009.4729 [hep-ph]}}].

\bibitem{Beuf:2011xd}
G.~Beuf, {\it {NLO} corrections for the dipole factorization of dis structure
  functions at low x},
  \href{http://dx.doi.org/10.1103/PhysRevD.85.034039}{{\em Phys. Rev.} {\bf
  D85} (2012) 034039} [\href{http://arXiv.org/abs/1112.4501}{{\tt
  arXiv:1112.4501 [hep-ph]}}].

\bibitem{Stasto:2013cha}
A.~M. Stasto, B.-W. Xiao and D.~Zaslavsky, {\it Towards the test of saturation
  physics beyond leading logarithm},
  \href{http://dx.doi.org/10.1103/PhysRevLett.112.012302}{{\em Phys. Rev.
  Lett.} {\bf 112} (2014) 012302} [\href{http://arXiv.org/abs/1307.4057}{{\tt
  arXiv:1307.4057 [hep-ph]}}].

\bibitem{Chirilli:2011km}
G.~A. Chirilli, B.-W. Xiao and F.~Yuan, {\it One-loop factorization for
  inclusive hadron production in {pA} collisions in the saturation formalism},
  \href{http://dx.doi.org/10.1103/PhysRevLett.108.122301}{{\em Phys. Rev.
  Lett.} {\bf 108} (2012) 122301} [\href{http://arXiv.org/abs/1112.1061}{{\tt
  arXiv:1112.1061 [hep-ph]}}].

\bibitem{Chirilli:2012jd}
G.~A. Chirilli, B.-W. Xiao and F.~Yuan, {\it Inclusive hadron productions in
  {pA} collisions},  \href{http://dx.doi.org/10.1103/PhysRevD.86.054005}{{\em
  Phys. Rev.} {\bf D86} (2012) 054005}
  [\href{http://arXiv.org/abs/1203.6139}{{\tt arXiv:1203.6139 [hep-ph]}}].

\bibitem{Stasto:2014sea}
A.~M. Sta{\'s}to, B.-W. Xiao, F.~Yuan and D.~Zaslavsky, {\it Matching collinear
  and small $x$ factorization calculations for inclusive hadron production in
  $pa$ collisions},  \href{http://dx.doi.org/10.1103/PhysRevD.90.014047}{{\em
  Phys. Rev.} {\bf D90} (2014)no.~1 014047}
  [\href{http://arXiv.org/abs/1405.6311}{{\tt arXiv:1405.6311 [hep-ph]}}].

\bibitem{Kang:2014lha}
Z.-B. Kang, I.~Vitev and H.~Xing, {\it Next-to-leading order forward hadron
  production in the small-$x$ regime: rapidity factorization},
  \href{http://dx.doi.org/10.1103/PhysRevLett.113.062002}{{\em Phys. Rev.
  Lett.} {\bf 113} (2014) 062002} [\href{http://arXiv.org/abs/1403.5221}{{\tt
  arXiv:1403.5221 [hep-ph]}}].

\bibitem{Altinoluk:2014eka}
T.~Altinoluk, N.~Armesto, G.~Beuf, A.~Kovner and M.~Lublinsky, {\it
  Single-inclusive particle production in proton-nucleus collisions at
  next-to-leading order in the hybrid formalism},
  \href{http://dx.doi.org/10.1103/PhysRevD.91.094016}{{\em Phys. Rev.} {\bf
  D91} (2015) 094016} [\href{http://arXiv.org/abs/1411.2869}{{\tt
  arXiv:1411.2869 [hep-ph]}}].

\bibitem{Watanabe:2015tja}
K.~Watanabe, B.-W. Xiao, F.~Yuan and D.~Zaslavsky, {\it Implementing the exact
  kinematical constraint in the saturation formalism},
  \href{http://arXiv.org/abs/1505.05183}{{\tt arXiv:1505.05183 [hep-ph]}}.

\bibitem{Dominguez:2011wm}
F.~Dominguez, C.~Marquet, B.-W. Xiao and F.~Yuan, {\it Universality of
  unintegrated gluon distributions at small x},
  \href{http://dx.doi.org/10.1103/PhysRevD.83.105005}{{\em Phys. Rev.} {\bf
  D83} (2011) 105005} [\href{http://arXiv.org/abs/1101.0715}{{\tt
  arXiv:1101.0715 [hep-ph]}}].

\bibitem{Kovchegov:2013cva}
Y.~V. Kovchegov and M.~D. Sievert, {\it {Sivers} function in the quasiclassical
  approximation},  \href{http://dx.doi.org/10.1103/PhysRevD.89.054035}{{\em
  Phys. Rev.} {\bf D89} (2014) 054035}
  [\href{http://arXiv.org/abs/1310.5028}{{\tt arXiv:1310.5028 [hep-ph]}}].

\bibitem{Balitsky:2015qba}
I.~Balitsky and A.~Tarasov, {\it Rapidity evolution of gluon {TMD} from low to
  moderate $x$},  \href{http://arXiv.org/abs/1505.02151}{{\tt arXiv:1505.02151
  [hep-ph]}}.

\bibitem{Kotko:2015ura}
P.~Kotko, K.~Kutak, C.~Marquet, E.~Petreska, S.~Sapeta and A.~van Hameren, {\it
  Improved {TMD} factorization for forward dijet production in dilute-dense
  hadronic collisions},  \href{http://arXiv.org/abs/1503.03421}{{\tt
  arXiv:1503.03421 [hep-ph]}}.

\bibitem{Kovchegov:2015zha}
Y.~V. Kovchegov and M.~D. Sievert, {\it Calculating {TMDs} of an unpolarized
  target: Quasi-classical approximation and quantum evolution},
  \href{http://arXiv.org/abs/1505.01176}{{\tt arXiv:1505.01176 [hep-ph]}}.

\bibitem{Metz:2011wb}
A.~Metz and J.~Zhou, {\it Distribution of linearly polarized gluons inside a
  large nucleus},  \href{http://dx.doi.org/10.1103/PhysRevD.84.051503}{{\em
  Phys. Rev.} {\bf D84} (2011) 051503}
  [\href{http://arXiv.org/abs/1105.1991}{{\tt arXiv:1105.1991 [hep-ph]}}].

\bibitem{Kovner:2010xk}
A.~Kovner and M.~Lublinsky, {\it Angular correlations in gluon production at
  high energy},  \href{http://dx.doi.org/10.1103/PhysRevD.83.034017}{{\em Phys.
  Rev.} {\bf D83} (2011) 034017} [\href{http://arXiv.org/abs/1012.3398}{{\tt
  arXiv:1012.3398 [hep-ph]}}].

\bibitem{Dumitru:2008wn}
A.~Dumitru, F.~Gelis, L.~McLerran and R.~Venugopalan, {\it Glasma flux tubes
  and the near side ridge phenomenon at {RHIC}},
  \href{http://dx.doi.org/10.1016/j.nuclphysa.2008.06.012}{{\em Nucl. Phys.}
  {\bf A810} (2008) 91} [\href{http://arXiv.org/abs/0804.3858}{{\tt
  arXiv:0804.3858 [hep-ph]}}].

\bibitem{Dumitru:2010iy}
A.~Dumitru, K.~Dusling, F.~Gelis, J.~Jalilian-Marian, T.~Lappi and
  R.~Venugopalan, {\it The ridge in proton-proton collisions at the {LHC}},
  \href{http://dx.doi.org/10.1016/j.physletb.2011.01.024}{{\em Phys. Lett.}
  {\bf B697} (2011) 21} [\href{http://arXiv.org/abs/1009.5295}{{\tt
  arXiv:1009.5295 [hep-ph]}}].

\bibitem{Dumitru:2014dra}
A.~Dumitru and A.~V. Giannini, {\it Initial state angular asymmetries in high
  energy {p+A} collisions: spontaneous breaking of rotational symmetry by a
  color electric field and {C-odd} fluctuations},
  \href{http://dx.doi.org/10.1016/j.nuclphysa.2014.10.037}{{\em Nucl. Phys.}
  {\bf A933} (2014) 212} [\href{http://arXiv.org/abs/1406.5781}{{\tt
  arXiv:1406.5781 [hep-ph]}}].

\bibitem{Lappi:2015vha}
T.~Lappi, {\it Azimuthal harmonics of color fields in a high energy nucleus},
  \href{http://dx.doi.org/10.1016/j.physletb.2015.04.015}{{\em Phys. Lett.}
  {\bf B744} (2015) 315} [\href{http://arXiv.org/abs/1501.05505}{{\tt
  arXiv:1501.05505 [hep-ph]}}].

\bibitem{Nikolaev:1991et}
N.~Nikolaev and B.~G. Zakharov, {\it Pomeron structure function and diffraction
  dissociation of virtual photons in perturbative {QCD}},
  \href{http://dx.doi.org/10.1007/BF01597573}{{\em Z. Phys.} {\bf C53} (1992)
  331}.

\bibitem{Kowalski:2003hm}
H.~Kowalski and D.~Teaney, {\it An impact parameter dipole saturation model},
  \href{http://dx.doi.org/10.1103/PhysRevD.68.114005}{{\em Phys. Rev.} {\bf
  D68} (2003) 114005} [\href{http://arXiv.org/abs/hep-ph/0304189}{{\tt
  arXiv:hep-ph/0304189}}].

\bibitem{Kowalski:2006hc}
H.~Kowalski, L.~Motyka and G.~Watt, {\it Exclusive diffractive processes at
  {HERA} within the dipole picture},
  \href{http://dx.doi.org/10.1103/PhysRevD.74.074016}{{\em Phys. Rev.} {\bf
  D74} (2006) 074016} [\href{http://arXiv.org/abs/hep-ph/0606272}{{\tt
  arXiv:hep-ph/0606272}}].

\bibitem{Kovalchuk:2014hea}
{\bf ZEUS} collaboration, N.~Kovalchuk, {\it {Charmonium production at HERA}},
  {\em PoS} {\bf DIS2014} (2014) 187.

\bibitem{Brodsky:1994kf}
S.~J. Brodsky, L.~Frankfurt, J.~Gunion, A.~H. Mueller and M.~Strikman, {\it
  Diffractive leptoproduction of vector mesons in {QCD}},
  \href{http://dx.doi.org/10.1103/PhysRevD.50.3134}{{\em Phys. Rev.} {\bf D50}
  (1994) 3134} [\href{http://arXiv.org/abs/hep-ph/9402283}{{\tt
  arXiv:hep-ph/9402283 [hep-ph]}}].

\bibitem{Abbas:2013oua}
{\bf ALICE} collaboration, E.~Abbas {\em et.~al.}, {\it Charmonium and $e^+e^-$
  pair photoproduction at mid-rapidity in ultra-peripheral {Pb-Pb} collisions
  at $\sqrt{s_{\rm NN}}$=2.76 {TeV}},
  \href{http://dx.doi.org/10.1140/epjc/s10052-013-2617-1}{{\em Eur. Phys. J.}
  {\bf C73} (2013) 2617} [\href{http://arXiv.org/abs/1305.1467}{{\tt
  arXiv:1305.1467 [nucl-ex]}}].

\bibitem{Lappi:2010dd}
T.~Lappi and H.~M{\"a}ntysaari, {\it Incoherent diffractive
  {$J/\psi$}-production in high energy nuclear {DIS}},
  \href{http://dx.doi.org/10.1103/PhysRevC.83.065202}{{\em Phys. Rev.} {\bf
  C83} (2011) 065202} [\href{http://arXiv.org/abs/1011.1988}{{\tt
  arXiv:1011.1988 [hep-ph]}}].

\bibitem{Toll:2012mb}
T.~Toll and T.~Ullrich, {\it Exclusive diffractive processes in electron-ion
  collisions},  \href{http://dx.doi.org/10.1103/PhysRevC.87.024913}{{\em Phys.
  Rev.} {\bf C87} (2013) 024913} [\href{http://arXiv.org/abs/1211.3048}{{\tt
  arXiv:1211.3048 [hep-ph]}}].

\bibitem{Adam:2015sia}
{\bf ALICE} collaboration, J.~Adam {\em et.~al.}, {\it Coherent {$\psi$(2S)}
  photo-production in ultra-peripheral {Pb-Pb} collisions at {$\sqrt{s_{\rm
  NN}}$ = 2.76 TeV}},  \href{http://arXiv.org/abs/1508.05076}{{\tt
  arXiv:1508.05076 [nucl-ex]}}.

\bibitem{Abelev:2012ba}
{\bf ALICE} collaboration, B.~Abelev {\em et.~al.}, {\it Coherent {$J/\psi$}
  photoproduction in ultra-peripheral {Pb-Pb} collisions at {$\sqrt{s_{NN}} =
  2.76$}~{TeV}},  \href{http://dx.doi.org/10.1016/j.physletb.2012.11.059}{{\em
  Phys. Lett.} {\bf B718} (2013) 1273}
  [\href{http://arXiv.org/abs/1209.3715}{{\tt arXiv:1209.3715 [nucl-ex]}}].

\bibitem{Aaij:2014iea}
{\bf LHCb} collaboration, R.~Aaij {\em et.~al.}, {\it Updated measurements of
  exclusive {$J/\psi$} and {$\psi$(2S)} production cross-sections in pp
  collisions at {$\sqrt{s}=7$~TeV}},
  \href{http://dx.doi.org/10.1088/0954-3899/41/5/055002}{{\em J. Phys.} {\bf
  G41} (2014) 055002} [\href{http://arXiv.org/abs/1401.3288}{{\tt
  arXiv:1401.3288 [hep-ex]}}].

\bibitem{CMS:2014ies}
{\bf CMS} collaboration, {\it Photoproduction of the coherent {$J/\psi$}
  accompanied by the forward neutron emission in ultra-peripheral {PbPb}
  collisions at {2.76~TeV}},  2014.
\newblock \href{http://cds.cern.ch/record/1971267}{CMS-PAS-HIN-12-009}.

\end{thebibliography}\endgroup
\bibliographystyle{JHEP-2modlong}

\end{document}